\documentclass{article}

\usepackage[utf8]{inputenc}
\usepackage[T1]{fontenc}
\usepackage[mathscr]{euscript}
\usepackage{geometry}
\usepackage{amsmath} 
\usepackage{amsfonts}
\usepackage{graphicx}
\usepackage{dcolumn}
\usepackage{upgreek}
\usepackage{slashed}
\usepackage{amssymb,amsfonts}
\usepackage{cancel}
\usepackage{bm} 
\usepackage{amsmath,mathrsfs} 
\usepackage{cite}
\usepackage{tikz}
\usetikzlibrary{positioning,shapes,fit,arrows}
\definecolor{myblue}{RGB}{56,94,141}
\definecolor{myred}{RGB}{30,94,100}

\def\beq{\begin{eqnarray}}
\def\eeq{\end{eqnarray}}

\catcode`\@=11
\@addtoreset{equation}{section}

\begin{document}

\title{\bf{NON-POLYNOMIAL LAGRANGIAN APPROACH \\ TO REGULAR BLACK HOLES}}

\author{Aimeric Coll\'{e}aux\footnote{e-mail: aimeric.colleaux@unitn.it}, Stefano Chinaglia\footnote{e-mail: s.chinaglia@unitn.it}, Sergio Zerbini\footnote{e-mail:sergio.zerbini@unitn.it} \\
\smallskip \\
\it Dipartimento di Fisica, Università di Trento and TIFPA-INFN \\
\it Via Sommarive 14, 38123 Trento, Italia}

\date{\today}

\maketitle

\begin{abstract}
We present a review on Lagrangian models admitting spherically symmetric regular black holes, and cosmological bounce solutions. Non-linear electrodynamics, non-polynomial gravity, and fluid approaches are explained in details. They consist respectively in a gauge invariant generalization of the Maxwell Lagrangian, in modifications of the Einstein-Hilbert action via non-polynomial curvature invariants, and finally in the reconstruction of density profiles able to cure the central singularity of black holes. The non-polynomial gravity curvature invariants have the special property to be second order and polynomial in the metric field, in spherically symmetric spacetimes. Along the way, other models and results are discussed, and some general properties that regular black holes should satisfy are mentioned. A covariant Sakharov criterion for the absence of singularities in dynamical spherically symmetric spacetimes is also proposed and checked for some examples of such regular metric fields.
\end{abstract}

\bigskip


\tableofcontents

\section*{Introduction}
\addcontentsline{toc}{section}{Introduction}

General Relativity (GR) has passed, up to now, every experimental tests. The last ones being the prediction of the wave-forms of gravitational waves coming from the merger of two black holes (BH) \cite{LIGO}, and two Neutron stars \cite{LIGO2}, detected by the LIGO-VIRGO collaboration. Since the first detection, many other BH-BH mergers have been detected, all of them in agreement with General Relativity \cite{LIGO3,LIGO4,LIGO5,LIGO6}.

However, this theory is also plagued by singularities \cite{SingTh1,SingTh2,SingTh3}, that indicate the breakdown of its predictivity at very small distances, and in the very early universe, i.e.  at the center of black holes and around the time of the big bang. Paradigmatic examples are the static Schwarzschild metric and its rotating and charged generalizations, as well as the GR solution with ordinary matter or radiation in a Friedmann-Lema\^itre-Robertson-Walker (FLRW) spacetime. 

Fortunately, such an ultraviolet catastrophe has already been encountered with the classical model of the hydrogen atom, where it was found that quantum mechanics solve the issue. For this reason, it is widely expected that a complete theory of quantum gravity should solve the singularity issue of General Relativity. 
\\

Assuming this to be true, two options are available to investigate the issue. Either there exists, like for the hydrogen atom \cite{SEMICLASS1,SEMICLASS2}, a semi-classical effective description of black holes and early times of the universe that are singularity free, or one needs a full non-perturbative theory of quantum gravity to resolve the singularity issue, in which case, it is possible that no effective metric descriptions would be available. 

Given that we do not have yet a consensual theory of quantum gravity, we will focus in this review on effective description of singularity free black holes, usually referred to as non-singular, or regular black holes (RBHs), and singularity free homogeneous and isotropic universes. Note however that there are already a number of proposals emerging directly from quantum gravity approaches (see for example  \cite{PLANCK,PLANCK2,LQG,LQG2,AsymSaf1,AsymSaf2,CONFORMAL1,CONFORMAL2,CONFORMAL3}).

The first attempt to regularize a black hole metric was done by Duan in 1954 \cite{Duan}, while the first example of RBH was proposed by Bardeen during the GR5 conference fourteen years later \cite{Bardeen} and since then many other examples have been provided. 

There are many possible roads, within an effective metric setting, to remove the singularity from Schwarzschild black hole. 
\\

First of all, one should choose a particular definition of singularity, either as geodesic incompleteness, in what case some effective wormhole metrics have been found to solve the problem \cite{WH1,WH2,WH3,WH4}, or one can focus on the regularity of curvature invariants, what leads to effective metric fields describing regular black holes. This is the approach we are considering here, and a non-exhaustive list of works on these RBHs includes \cite{scale,Is, M, Hayward_1, Bronnikov, Elizalde, Dymnikova, GSP, Dymnikova92, NSS, ANSS, Modesto, CMSZ, Dymnikova_2, Culetu, Maeda, Horava, Horava_1, KS, CCO, Pradhan, Ma, Johannsen, Rodrigues, Fan, Beato, Frolov_1, Frolov_2, Frolov_3, Frolov_4, Frolov_5, Frolov_6, Frolov_7, Vag1, Vag2, CZ, NO17, ans,fuzzi}. Each of these papers deals with an asymptotic Schwarzschild behavior, either when the observer is far from the source, or when the deformation parameters are vanishing, while at the centre, a (anti) de Sitter core is present, replacing the singularity. This is a sufficient condition to regularize the metric, known as Sakharov criterion \cite{Sakharov}: namely the stress energy tensor reads $T_{\mu\nu} \simeq \Lambda_0 \, g_{\mu\nu}$ near the origin, where $\Lambda_0$ is an effective cosmological constant. 
Note that, as shown in \cite{WH3}, there exists some particular metric fields where these two definitions are incompatible, namely, where the spacetime is geodesically complete, but where invariants are singular, and reciprocally. 

In the Appendix, a generalized covariant criterion is formulated, with a brief review on the spacetimes singularity issues. This criterion works both for static and dynamical spherically symmetric spacetimes, and is related to the so-called Limiting Curvature Conjecture \cite{M,Mar,Frolov_6}. 

For the sake of completeness, remark that these two ways of removing the singularity by static regular metric fields are not necessary. Indeed, one can also study directly gravitational collapse \cite{Stellar,Barcelo,Malafarina,Bambi1,Bambi2,Bambi3,Bambi4,Casadio}, with quantum-like corrections either in the matter sector, or in the curvature one. In both cases, the resulting singularity-free spacetime might be different from either static wormholes, or static regular black holes with (A)dS cores, for example, it might describe a dynamical bounce. In this view, static RBHs might be only a particular regime of these more physically complete spacetimes. 

In all these cases, a side-effect of removing the singularity is expected to be also a dissolution of the information loss paradox (see \cite{INFOLOSS,INFOLOSS2,INFOLOSS3} and references therein) : if matter is not destroyed by the singularity inside a black hole, either it can directly goes out (after a long time for the external observer) via a bounce \cite{Bambi1,Bambi2,Bambi3,Bambi4,Casadio,PLANCK}, or via evaporation \cite{Hayward_1,Frolov_4} or finally be locked inside in the form of a regular stable extremal remnant \cite{INFOLOSS3,REMN1,REMN2,REMN3,REMN4,REMN5,REMN6}, that might wait for a non-perturbative effect to release its matter content. Note that this last possibility might also be a candidate for dark matter \cite{REMN6,REMN7,REMN8}.
\\

In this review, we will focus on corrections to General Relativity arising from Lagrangian based models, in Section 1. and 2., and from an effective fluid description, in Section 3., that can also be seen as coming from an unknown Lagrangian. Other approaches not based on a Lagrangian to deal with singularities are for example \cite{OTHER0,OTHER1,OTHER2,OTHER3,OTHER4,OTHER5,OTHER6}.

There are three main possibilities to modify the action of General Relativity (that are not mutually incompatible), and RBHs have been found in all of these : either one modifies the matter sector, in the form of a minimal \cite{Elizalde}, or non-minimal coupling with gravity \cite{Dereli,Sert,Balakin,Zayats,BLZ,Horndeski,Drummond,NMYM2}
, or one modifies directly the geometrical sector. Usually, non-minimally coupled additional fields can be seen either as matter fields or as new gravitational degrees of freedom, depending on the context, and sometimes, depending only on the conformal frame we are looking in, the paradigmatic example being $F(R)$ gravity \cite{DEFELICE}.

The first one, that we study in Section 1 and Section 3 has been followed by the majority of the papers, and it involves a modification of the stress-energy tensor in the form of exotic matter. It is indeed the simplest one, although it has to deal with a major problem. Indeed, the general solution of such approach typically contains a singular term $C/r$, coming from the integration of the Einstein Equations. It is then possible to achieve a regular black hole, as soon as one is able to use a suitable physical boundary condition in order to rid off this term. With regard to this issue, see \cite{Nicolini}. In our previous paper \cite{CZ}, we argued that the presence of an integration constant $C$ is a mathematical fact that cannot be avoided; a possible way to deal with this issue is presented in \cite{C2017}. Of course, it remains possible to require the regularity of the solution in order to get rid of this singular term. 

In Section 1. we will see a modification of the matter part, in the form of Non-linear Electrodynamics (NED) Lagrangians minimally coupled to gravity \cite{Bronnikov, Dymnikova, Dymnikova_2, Culetu, Pradhan}. This particular approach has been developed in the late '90 by E. Ayon-Beato and A. Garcia. They proposed a scheme to generate (even regular) solutions from a NED Lagrangian minimally coupled to the standard Lagrangian of GR \cite{Beato}. They successfully applied their scheme to build up a modified Reissner-Nordstr\"{o}m, Bardeen-like, solution, later identified as a magnetic monopole solution \cite{Beato_1}. Their scheme was later generalized by I. Dymnikova \cite{Dymnikova} for spherical and static solutions. Further properties can be found in \cite{NBS}. Unfortunately, this approach is very difficult to be implemented in practice and very rarely, at Lagrangian level, one gets full analytic results. For this reason, the so called dual P approach \cite{GSP} has been proposed, in which central role is not played by the NED Lagrangian.
\\

In Section 2. we will focus on corrections to the geometrical sector able to find regular black holes without specific and regular matter distributions, i.e. able to cure the singularity arising from a point-like mass \cite{NSS, ANSS, Modesto, Maeda}. More precisely, we will deal with Non-polynomial gravity (NPG), see \cite{Deser} for the origin of the approach, and \cite{RBHNPG} for RBHs.

This approach can be seen as a generalization of Lovelock-Lanczos gravity \cite{LLG1,Lovelock_2} and Quasi-Topological gravity \cite{QTG1,QTG2,QTG3,QTG4,QTG5,QTG6,QTG7,QTG8}, in the sense that the polynomiality and $2$sd order equations of motion are required only for specific spacetimes, here, spherically symmetric or cosmological ones. Therefore, the curvature invariants involved in the NPG corrections are non-polynomial for general metric fields, but nonetheless polynomial in spherically symmetric spacetimes. We will see by reviewing the results of \cite{RBHNPG}, that a minimal second order generalization of Einstein gravity in spherical symmetry leads quite simply to RBHs solutions, and that the first and second corrections to Einstein-Hilbert action give respectively the Poisson-Israel \cite{PI} and Non-minimal Einstein-Yang-Mills (NMYM) \cite{BLZ,NMYM2} regular black hole solutions. This last corresponds to a non-minimal coupling to gravity of a $SU(2)$ Y-M field in a Wu-Yang monopole configuration \cite{Zayats}.

In this section, we will also review some basic properties that a physical regular black hole is expected to satisfy. To deal with these constraints, a NPG unified approach and its relation with generic two dimensionally reduced models \cite{Maeda,TavesK,Kunstatter,Designer,Dilaton1} are reviewed. It consists in a 4D generalization of the very general 2D dilaton gravity model, presented in \cite{Maeda}, where a subclass was the so-called Lovelock Designer models, a generalization of Lovelock-Lanczos gravity in spherically symmetric spacetimes, able to produce quite simply regular black holes. Finally, we will also present a cosmological NPG model, found in \cite{CCZ17}, that is a generalization of Lovelock-Lanczos gravity in FLRW, and that is able to reproduce the loop quantum cosmology bounce solution \cite{Ashtekar,Ashtekar_2,bo,Bojo}. See also \cite{Tu1, Tsu, add, No, Bra, od, B, lo, staro, Gie, I, qiu, Dz, Malkiewicz, Malkiewicz_2, bo, Ashtekar, Ashtekar_2, Helling, Helling_2, Cai, Cai2} for other approaches able to cure the big bang singularity.

\section{Non Linear Electrodynamics}

We begin our Lagrangian study of regular black holes by presenting a popular and well-known approach to find regular black holes : Non-linear electrodynamics. It enters in the class of models that consider matter fields minimally coupled to Einstein gravity as a possible solution of the singularity issue. In this case, the matter field is an electromagnetic one, with a generalized gauge invariant Lagrangian. This approach  has been  discussed in several papers, see, for example \cite{GSP,Beato,Dymnikova,breton,NBS}.

The NED gravitational model is based on the following action 
\begin{equation}
\label{action}
\mathscr{I} = \int d^4 x \ \sqrt{-g} \left(\frac{R}{2} - 2\Lambda -\mathscr{L}(I)\right)\,,
\end{equation}
where $R$ is the Ricci scalar, $\Lambda$ is a cosmological constant, and  $I=\frac{1}{4}F^{\mu\nu} F_{\mu\nu}$ is an electromagnetic-like tensor and $\mathscr{L}(I)$ is a suitable function of it. Recall that $F_{\mu\nu}=\partial_\mu A_\nu - \partial_\nu A_\mu$. We will only deal with gauge invariant quantities, and  we put $\Lambda=0$, because its contribution can be easily  restored. The equations of motion read
\begin{equation}
\label{eq1}
G^\nu_\mu  = -F_{\alpha\mu} \partial_I \mathscr{L} F^{\nu\alpha} + \mathscr{L}\delta^\nu_\mu 
\end{equation}
 \begin{equation}
\label{eq2}
\nabla^\mu (F_{\mu\nu} \partial_I \mathscr{L}) = 0\,.
\end{equation}

Another equivalent approach is called dual P approach and it is based on   two  new  gauge invariant quantities \cite{GSP}
\begin{equation}
\label{P}
P_{\mu\nu}\equiv F_{\mu\nu} (\partial_I \mathscr{L}(I))\,,\,\, P \equiv \frac{1}{4} P_{\mu\nu}P^{\mu\nu}\,,
\end{equation}
such that $\nabla^\mu P_{\mu\nu} = 0$.

In the following, we shall make use only of the traditional approach based on equations (\ref{eq1}) and (\ref{eq2}). Within the static spherically symmetric ansatz, given by 
\begin{equation}
\label{metric0}
ds^2 = -f(r)dt^2 + \frac{1}{g(r)} dr^2 + r^2 d\Omega^2\,,
\end{equation}
and from  (\ref{eq2}), one has 
\begin{equation}
\label{4}
\partial_r\left(r^2 \partial_I \mathscr{L}F^{0r}\right)   = 0\,,
\end{equation}
We will consider in this section, $g(r)=f(r)$. Since $I=\frac{1}{2}F_{0r}F^{0r}=- \frac{1}{2}F_{0r}^2$, one gets
\begin{equation}
\label{r111}
r^2 \partial_I \mathscr{L}   = \frac{Q}{\sqrt{-2 I}}\,,
\end{equation}
$Q$ being a constant of integration. As a result, within this  NED approach,  one may solve the generalized Maxwell equation. We shall make use of this equation and the (t,t) component of the Einstein equation, which reads
\begin{equation}
\label{r}
G^t_t  =\frac{rf'+f-1}{r^2} = 8\pi\left( -2 I\partial_I \mathscr{L}+ \mathscr{L}\right)=-8\pi \rho\,.
\end{equation}
Introducing the  more convenient quantity  $X \equiv Q\sqrt{-2  I}$, one may rewrite equation (\ref{r111}) as
\begin{equation}
\label{r2}
r^2 \partial_X \mathscr{L}   = 1\,.
\end{equation}
Furthermore, we have
\begin{equation}
\label{r4}
 \rho= X\partial_X \mathscr{L} - \mathscr{L}=\frac{X}{r^2}-\mathscr{L}  \,.
\end{equation}
Thus, when  $ \mathscr{L}(X) $ is given, then making use of (\ref{r111}), one may obtain $\rho=\rho(r)$. This allows to discuss the problem even within the formalism of a (perfect) fluid (see the last section).

\subsection{Reconstruction}

Here we present a discussion on a quite efficient reconstruction scheme able to produce RBH solutions. Within the NED framework, the  Einstein equation, re-written  as 
\begin{equation}
\label{ege55}
\frac{d}{dr}\left( r(f-1)\right) = - 8\pi r^2 \rho\,,
\end{equation}
gives $\rho(r)$ once $f(r)$ is known.  The other two equations are
\begin{equation}
\label{r_1}
r^2 \partial_X \mathscr{L} =1 \,, \quad \mathscr{L}=\frac{X}{r^2}-\rho\,.
\end{equation}
From the above equations,  we get $ X = - r^3 \rho'/2$, the prime indicating the radial derivative. As a consequence, one may obtain $r=r(X)$, and, making use of the second equation,   $\mathscr{L}=\mathscr{L}(X)$.

As warm up, let us start from the singular Reissner-Nordstr\"{o}m  solution
\begin{equation}
\label{SdSRN_solution}
f(r) = 1-\frac{C}{r} + \frac{Q^2}{r^2} 
\end{equation}
Making use of (\ref{ege55}), one has $\rho=\frac{Q^2}{8\pi r^4 }$ and it is easy to show that the Maxwell Lagrangian is recovered. 

As a second less trivial example, let us consider the general solution 
\begin{equation}
\label{f_example_1}
f(r) = 1 -\frac{C}{r} -\frac{2A}{\xi} + \frac{2A}{\xi} \frac{\arctan{\left(\frac{r}{\xi}\right)}}{r} -H_0^2r^2\,,
\end{equation}
where $C$ is an integration constant and $A$, $\xi$ and $H_0$ are suitable parameters. This solution is a generalization of  black hole solution obtained from a particular Horndeski Lagrangian, namely Einstein gravity with a non minimally coupled scalar field \cite{MR}.

The effective density can be obtained by (\ref{ege55}), namely
\begin{equation}
\label{r-3}
\rho = \frac{A}{4\pi \xi^2(\xi^2 + r^2)}+\frac{3 H_0^2}{8\pi}\,.
\end{equation} 
Thus
\begin{equation}
\label{r-4}
X=\frac{B r^4}{(\xi^2 + r^2)^2}\,, \quad \text{where} \quad B=\frac{A}{4\pi}>0\,.
\end{equation} 
The Lagrangian can easily  be  reconstruct, and the result is
\begin{equation}
\label{chin1}
 \mathscr{L}(X)=-\frac{1}{\xi^2}\left(\sqrt{B}-\sqrt{ X} \right)^2-\frac{3H^2_0}{8\pi} \,.
\end{equation}
From eq. (\ref{r-4}), one sees that suitable choice for the parameters $A$ and $H_0$ (e.g. $A, \ H_0 \geq 0$) is able to satisfy the Weak Energy condition (WEC). Indeed we recall that  the WEC is satisfied if and only if \cite{Hawking_Ellis}
\begin{align}
\label{WEC}
& \rho \geq 0 \\
& \rho + p_k \geq 0 \ \ \ \ \ k=1,2,3\,.
\end{align}
The issue if the WEC is satisfied or not by regular and in particular by NED regular solutions has been widely discussed, among others, by I. Dymnikova in \cite{Dymnikova_WEC_1} and \cite{Dymnikova_WEC_2}. In particular, \cite{Dymnikova_WEC_1} finds some conditions a Lagrangian must satisfy, in order to fulfill the WEC. We also recall that the Dominant Energy Condition (DEC) \cite{Hawking_Ellis} is given by the WEC plus the additional condition $\rho-p_k \geq 0$. 

As a last example, let us start from the following metric, 
\begin{equation}
\label{f_2}
f(r) = 1 -\frac{C}{r} +\frac{4B\pi}{r^2+\xi^2} - \frac{4B\pi}{\xi} \frac{\arctan{\left(\frac{r}{\xi}\right)}}{r}\,.
\end{equation}
with $B >0$. If $C=0$, this is the regular black hole solution proposed by Dymnikova in \cite{Dymnikova_1}. The solution is asymptotically flat, and its regular part has de Sitter core and  no conical singularity. Let us try to reconstruct the related NED Lagrangian.
Again, from (\ref{ege55}), one has
\begin{equation}
\label{r-5}
\rho = \frac{B}{(\xi^2 + r^2)^2}\,,
\end{equation} 
and since $B$ is positive, this density clearly satisfies the WEC. In this case, also DEC is
satisfied. Furthermore,
\begin{equation}
\label{r-6}
X = \frac{2 B r^4}{(\xi^2 + r^2)^3}\,.
\end{equation}
We may re-write it as
\begin{equation}
\label{r-7}
X(\xi^2 + r^2)^3=2B (r^2)^2\,,
\end{equation} 
and consider  $r^2$ as a function of $X$, obtained solving an algebraic equation of third order. Once we  have the solution,  the Lagrangian reads
\begin{equation}
\label{r-8}
\mathscr{L}(X) = \frac{(r^2(X) -\xi^2) X}{ 2(r^2(X))^2}\,. 
\end{equation}
The final expression can be written explicitly, but it is too complicate, and it will not be written here.  However we are able to write $\mathscr{L}$ with the simple parametric representation
\begin{equation}
\label{r-9}
\mathscr{L}(r) = \frac{(r^2 -\xi^2)}{(r^2+\xi^2)^3}\,,
\end{equation}

Finally we conclude this Section with the following remark. One can start with the $\mathscr{L}$ given in an implicit form $X=G(\mathscr{L})$, where $G$ is a smooth known  function. Taking the derivative with respect to $X$, and making use of  
$r^2 \partial_X\mathscr{L}=1$, one has $r^2=\partial_{\mathscr{L}}G(\mathscr{L})$. In principle, this gives $\mathscr{L}$ as function of $r$, and the effective density may be computed $r^2 \rho =G(\mathscr{L})-r^2  \mathscr{L}$.

For example, let us consider
\begin{equation}
\label{r14}
X=G(\mathscr{L})=G_0+G_1\mathscr{L}+\frac{G_2}{2} \mathscr{L}^2 \,,
\end{equation}
with $G_{0, 1, 2}$ suitable constants. Then one has
\begin{equation}
\mathscr{L}=\frac{r^2-G_1}{G_2}\, \quad \text{and} \quad r^2 \rho =G_0-\frac{G_1^2}{2G_2}+r^2 \frac{G_1}{G_2}-\frac{r^4}{2G_2}  \,. 
\end{equation}
In order to avoid the presence of conical singularities, one has the constraint $G_0 = G_1^2/2G_2$ (see e.g. \cite{Solodukhin} and references therein). As a consequence, the Lagrangian is determined by the algebraic equation
\begin{equation}
\label{r14a}
X=\frac{G_1^2}{2G_2} +G_1\mathscr{L}+\frac{G_2}{2} \mathscr{L}^2 \,,
\end{equation}
and one gets
\begin{equation}
\label{r13ab}
r^2 \rho =r^2 \frac{G_1}{G_2}-\frac{r^4}{2G_2}  \,.
\end{equation}
The related general solution reads
\begin{equation}
\label{f_example_6}
f(r) = 1 -\frac{C}{r}  - 8\pi \left( \frac{r^2 G_1}{3G_2}-\frac{r^3}{10G_2}   \right) \,.
\end{equation}
To our knowledge, this is a new static spherically symmetric solution. Other solutions can be found with the same technique,  the above solution being the simplest one. Moreover, if $G_1$ and $G_2$ are negative, the WEC is satisfied. However, since $f(r)$ contains the cubic term $r^3$, for large $r$, the Kretschmann scalar will diverges like $r^2$. Thus, it is singular at infinity. It is worth to mention that the covariant version of the Sakharov criterion, that we will discuss in Appendix A, requires the scalar $\Phi(r)=\frac{1-f(r)}{r^2}$ to be bounded for every $r$: this implies that asymptotically $f(r)$ should go no faster than $r^2$ for large $r$.

A final remark is in order. Throughout this section, we considered only theories generated by actions of the kind of action (\ref{action}). In principle, one can use also more complicated actions, as is done in \cite{Junior}. However this strategy does not appear to be able to cancel the singularity issue in NED, since it is related to the electromagnetic part of the theory $-$ which remains invariant.

\section{Non-polynomial gravity}

Now that we have seen a matter-like modification to the Einstein equations that is able to produce regular black holes, we will, in this section, summarize the work of \cite{RBHNPG}, specializing in $4$ dimensions. The details about this approach can be found there. The idea is to extend the results of \cite{Deser,gao,CCZ17,ZerbAim} in which non-polynomial curvature invariants were considered, both in the context of black holes and cosmology, with the motivation of finding effective-like actions in these sectors. By this, we always mean polynomial actions (at least) in the considered class of spacetimes. We will see that lot's of regular black holes can be found via this approach, and that it provides a unified higher dimensional (4D in our case) description of $2$D dilaton gravity reduction models, studied for example in \cite{Maeda} and \cite{TavesK}. The bounce solution of loop quantum cosmology can also be found via this approach, as we will explain at the end of this section.

\subsection{Cotton tensor decomposition : order-$0$ curvature tensor}

We first introduce the Cotton tensor in four dimensions, defined by : 
\begin{eqnarray}
C_{\alpha\beta\gamma} = \nabla_\alpha R_{\beta \gamma} -  \nabla_\beta R_{\alpha \gamma} + \frac{1}{6} \left( g_{\alpha \gamma} \nabla_\beta R - g_{\beta \gamma} \nabla_\alpha R \right).
\end{eqnarray} 
In spherically symmetric spacetimes defined by the interval : 
\begin{eqnarray}
\label{metric}
ds^2 = d\Sigma^2 + d\Omega^2_{r} \, , 
\end{eqnarray}
where $d\Sigma^2 =\gamma_{AB} dx^A dx^B$, $A=\{1,2\}$, and $d\Omega^2_{r}=r^2 \left( d\theta^2 + \sin^2\theta \, d\phi^2 \right)$, the Cotton tensor obeys an algebraic equation :
\begin{eqnarray}
\label{alge}
\Big( -3 C^{\mu \alpha\beta} + C^{\mu \beta \alpha} \Big) C^{\nu}_{\;  \alpha\beta} =\frac{1}{2} \left( - 2 \, g^{\mu\nu} + \frac{3}{2} \sigma^{\mu\nu} \right) C^{\rho\sigma\gamma}C_{\rho\sigma\gamma} \, ,
\end{eqnarray} 
providing that $\sigma_{\mu\nu}$ is the four dimensional degenerate metric of the $2$-sphere $\Omega_{r}$ of radius $r$. Conversely, defining the following non-polynomial tensor
\begin{eqnarray}
u_{\alpha\beta} :=   \frac{\big( -3 C_{\alpha\mu\nu}+C_{\alpha\nu\mu} \big) C_{\beta}^{\;\;\mu\nu} }{C_{\sigma\rho\delta} C^{\sigma\rho\delta}} \; , 
\end{eqnarray}
allows to construct this degenerate metric of $\Omega_{r}$ in the following way : $\sigma_{\alpha\beta}  := \frac{4}{3}\left(  u_{\alpha\beta} + g_{\alpha\beta} \right)$. Note that this construction holds in any dimensions $d>3$ (see \cite{RBHNPG}).

In spacetimes (\ref{metric}), $u_{\alpha\beta}$ does not depend on the derivatives of the metric, and thus it is possible to construct curvature scalars from it that are polynomial and second order in (\ref{metric}), and thus, to construct effective-like actions, in a similar way as Lovelock-Lanczos gravity (that is polynomial and second order for any metric) \cite{LLG1,Lovelock_2}, or Quasi-Topological gravity (that is polynomial for any metric but second order only for spherical symmetry) (see \cite{QTG1,QTG2,QTG3,QTG4,QTG5,QTG6,QTG7,QTG8}) for details about these theories). If one sees QTG as a generalization of LLG, then NPG can be thought of as a generalization of this last, in the sense that this approach weakens the background independent polynomiality assumption. Moreover, the origin of Quasi-Topological gravity was found to be a very similar property as (\ref{alge}), but involving the Weyl tensor (see \cite{QTG1,Deser2}). Therefore, this new algebraic equation might also give rise to new QTGs.

Note also that, if we are interested by effective-like actions, where the invariants have a definite order, then, from this ``order-$0$" tensor, one can build a large but finite number of new scalars for each order. In this sense, the models and solutions built from property (\ref{alge}) are not reconstructed, but found. For example order-$2$ scalars are $\nabla_\gamma u_{\alpha\beta} \nabla^\gamma u^{\alpha\beta} $, $R^{\alpha\beta} u_{\alpha\beta}$,  $\nabla^\alpha \nabla^\beta u_{\alpha\beta}$, etc... For each order, we could sum them, and study systematically the solutions. However, we will use here a shortcut : we want an effective-like action that is a generalization of the Einstein-Hilbert one in spherically symmetric spacetimes. 

\subsection{Action}

 In order to do so, recall that in the class of spacetimes (\ref{metric}), $r(x)$ is a scalar field on $\Sigma$. Denoting with tildes the covariant derivatives of this manifold, we can write the following well-known $2$-dimensional decomposition of the Ricci scalar of the 4-dimensional manifold in (\ref{metric}) (see for example \cite{AshHay}) :
\begin{eqnarray}
 R \bigr\rfloor = R(\gamma) + R(\Omega) +2 \left( \frac{\tilde{\nabla} r .  \tilde{\nabla} r - \tilde{\Box} r^2 }{r^2}  \right) \, , 
\end{eqnarray}
where $R(\gamma)$ and $R(\Omega)$ are the Ricci scalars of the $2$-dimensional manifolds $\Sigma$ and $\Omega_r$, respectively, and we write for any scalar $X$, its restriction to (\ref{metric}) by $X\bigr\rfloor$. The idea of this approach is to build a series of higher order corrections to Einstein-Hilbert action from $u_{\alpha\beta}$ that keeps for any order the structure of the Ricci scalar $R$ and of the Einstein tensor $G_{\mu\nu}$. To do so, we consider the following action : 
\begin{eqnarray}
\label{NPGaction}
\mathcal{I} =\frac{1 }{16\pi G }  \int_\mathcal{M} d^{4}x  \, \sqrt{-g} \left( R -2 \Lambda + \sum _{i=1}^m l^{ i} \,  \mathcal{R}^{i/2} \, \Big(\alpha_i \,  \mathcal{R}+ \beta_i  \, \mathcal{S}_i \Big) \right) \, , 
\end{eqnarray}
where $l$ is a length scale introduced for dimensional reasons, and the scalars $\mathcal{R}$ and $\mathcal{S}_i$ are defined in terms of our non-polynomial tensor $u_{\alpha\beta}$ by
\begin{eqnarray}
\label{NPGscalarS}
\mathcal{S}_i = \frac{1}{3(i-2)} \left( - \frac{4 i (i+1) }{3} \nabla_\gamma u_{\alpha\beta} \nabla^\gamma u^{\alpha\beta}   + R + 4 \left(R^{\alpha\beta} - \nabla^\alpha \nabla^\beta \right) u_{\alpha\beta} \right) \, , 
\end{eqnarray}
and
\begin{eqnarray}
\label{NPGscalarR}
\mathcal{R} = \frac{2}{3}\left(  R +  \left(R^{\alpha\beta} - \nabla^\alpha \nabla^\beta \right) u_{\alpha\beta} \right) \, , 
\end{eqnarray}
in such a way that on spacetimes (\ref{metric}) they are : 
\begin{eqnarray}
\label{NPGscalars}
\mathcal{S}_i \bigr\rfloor = \frac{1}{2-i} \left( R(\gamma) + \frac{ (i^2 + i +2)\tilde{\nabla} r .  \tilde{\nabla} r - 2 \tilde{\Box} r^2  }{r^2} \right) \, ,\, \,\, \text{and} \,\,\,\, \mathcal{R} \bigr\rfloor =\frac{R(\Omega)}{2} \, .
\end{eqnarray}
Therefore, in spacetimes (\ref{metric}), the higher order corrections in action (\ref{NPGaction}) can be interpreted as a sum of powers of the Ricci scalar of the horizon manifold, given by $\mathcal{R}$, multiplied by a suitable scalar $\mathcal{S}_i$, which provides the same structure as General Relativity in (\ref{metric}) for any order of corrections $i$. The precise expression of the coefficients of the scalars $\mathcal{S}_i$ is explained in details in \cite{RBHNPG}, and is based on $3$ requirements : 1. that the equations of motion have the same structure as the Einstein tensor in spherical symmetry (see equation (\ref{2DcovEOM})), 2. the variational problem is well-posed, 3. the order-$d$ scalar (here $d=4$) $\mathcal{R} ^{(d-2)/2} \, \mathcal{S}_{d-2}$ contributes to the equations of motion (see equation (\ref{Topology})). After deriving the spherically symmetric solutions of these models, we will make a brief comment on this third point, which is related to the normalization $\frac{1}{2-i}$ that we used in the definition of $\mathcal{S}_i$.

From the definitions (\ref{NPGscalarS}) and (\ref{NPGscalarR}), we see that 
\begin{eqnarray}
R=2 \left( \mathcal{R} + \mathcal{S}_0 \right) .
\end{eqnarray}
Therefore, providing that  $\alpha_{-2} =-2  l^{2} \, \Lambda $, $\beta_{-2}=0$, $\alpha_{-1}=\beta_{-1}=0$ and $\alpha_{0}=\beta_0=2$, our action can be rewritten for any spacetimes as 
\begin{eqnarray}
\label{NPGaction2}
\mathcal{I} =\frac{1 }{16\pi G }  \int_\mathcal{M} d^{4}x  \, \sqrt{-g}  \sum _{i=-2}^m l^{ i} \,  \mathcal{R}^{i/2} \, \Big(\alpha_i \,  \mathcal{R}+ \beta_i  \, \mathcal{S}_i \Big) \, , 
\end{eqnarray}
from which the equations of motion and solutions can be written in a very compact form.

\subsection{Covariant $2$-dimensional Equations of motion}

Using (\ref{NPGaction2}) and (\ref{NPGscalars}), and noting that in spacetimes (\ref{metric}), the determinant of the metric splits into $\sqrt{-g}= \sqrt{-\gamma} \sqrt{\sigma} \,  r^2 $, we can integrate out the angular part of the action, what gives, discarding boundary terms, the dimensionally reduced action : 
\begin{eqnarray}
\mathcal{I}= \frac{\mathcal{A}_{2,1}}{16 \pi  G} \, \mathcal{I}_{\,2D} \, , 
\end{eqnarray}
with 
\begin{eqnarray}
 \mathcal{I}_{\,2D} =\sum _{i=-2}^m l^{ i} \int_\Sigma d^{2}x \, \sqrt{-\gamma}  \,  \left( \alpha_i  + \beta_i \left[ (1-i) \tilde{\nabla} r . \tilde{\nabla} r +  \frac{r^2}{2-i}  \, R(\gamma)  \right] \right)r^{-i}  \, ,
\end{eqnarray}
where $\mathcal{A}_{2,1} = \int d^2 x \sqrt{\sigma} = 4 \pi$ is the volume of the $2$-dimensional sphere of radius unity $\Omega_1$ and $\sigma$ is the determinant of the $2$-dimensional metric on this space.

The variation of $\mathcal{I}_{\,2D}$ with respect to the metric $\gamma_{AB}$ of $\Sigma$ and to the scalar field $r(x)$ on $\Sigma$ gives : 
\begin{eqnarray}
\begin{split}
\delta \mathcal{I}_{\,2D}  = \sum _{i=-2}^m l^{ i}  \int_\Sigma d^{2}x \,  \sqrt{-\gamma} \, &\Bigg\{   \bigg(-i \,  \alpha_i  +\beta_i \left(i (1-i) \tilde{\nabla} r . \tilde{\nabla} r +  r^2 R(\gamma) - 2 (1-i) r \tilde{\Box} r \right) \bigg) \delta r 
\\
+  r \bigg(-\frac{1}{2} \alpha_i  \, \gamma_{AB} - \beta_i \, r &\left( \tilde{\nabla}_A \tilde{\nabla}_B - \gamma_{AB} \tilde{\Box} \right) r 
+ \frac{\beta_i}{2} (1-i) \gamma_{AB} \tilde{\nabla} r . \tilde{\nabla} r   \bigg) \delta \gamma^{AB} \Bigg\}  \, r^{-i-1} 
\end{split}
\end{eqnarray}
from which follows the equations of motion given by the $4$-dimensional curvature tensor $\mathcal{E}_{\mu \nu}$ and the Stress-energy tensor $T_{\mu\nu}$ obeying the symmetry (\ref{metric}) :
\begin{eqnarray}
\begin{split}
\label{2DcovEOM}
&\mathcal{E}_{AB} := \sum _{i=-2}^m l^{ i}  \Big(-\frac{1}{2} \alpha_i  \, \gamma_{AB} - \beta_i \, r \left( \tilde{\nabla}_A \tilde{\nabla}_B - \gamma_{AB} \tilde{\Box} \right) r 
+ \frac{\beta_i}{2} (1-i) \gamma_{AB} \tilde{\nabla} r . \tilde{\nabla} r   \Big) r^{-i} = 8 \pi T_{AB} \, \\
\, &
\\
&\mathcal{E}_{\rho}^{\rho} = \mathcal{E}_{\phi}^{\phi} :=  \frac{1}{4} \sum _{i=-2}^m l^{ i}   \left( i \,  \alpha_i \,  -\beta_i \left(i (1-i) \tilde{\nabla} r . \tilde{\nabla} r +  r^2 R(\gamma) - 2 (1-i) r \tilde{\Box} r \right) \right)r^{-i-2} = 8 \pi T_{\rho}^{\rho}
\end{split}
\end{eqnarray}
The other components of $T_{\mu\nu}$ and $\mathcal{E}_{\mu \nu}$ are vanishing.
\\

In order to find the vacuum solutions of these equations of motion, the simplest way is to use the Weyl method, what is possible due to the principle of symmetric criticality \cite{PSC1,PSC2} applied to spherically symmetric spacetimes.

Going back to the 4 dimensional case, the action is : 
\begin{eqnarray}
\mathcal{I} = \frac{1 }{4 G} \int_\Sigma d^2x \, \sqrt{-\gamma} \, r^2    \sum _{i=-2}^m l^{ i} \,  \mathcal{R}^{i/2} \, \Big(\alpha_i \,  \mathcal{R}+ \beta_i  \, \mathcal{S}_i \Big) 
\end{eqnarray}
To apply the Weyl method, we follow closely \cite{DeserFrank}, where it was applied to GR, and choose the following gauge : 
\begin{eqnarray}
\label{GAUGE}
d\Sigma^{\,2} = -a(t,r) b(t,r)^2 dt^2 + \frac{dr^2}{a(t,r)} + 2 b(t,r) f(t,r) dt \, dr ,
\end{eqnarray}
for which the $\mathcal{S}$-term of the action gives (discarding boundary terms) :
\begin{eqnarray}
\sqrt{-\gamma} \, r^2\, \mathcal{R}^{i/2} \, \mathcal{S}_i \, \Bigr\rfloor = \frac{r^{1-i}}{\sqrt{1+f^2}}  \left( a b'-  f \frac{\dot{a}}{a} + \frac{a b f f'}{1+f^2}\right) \;,
\end{eqnarray}
where dots and primes are derivatives with respect to time and radius respectively. It makes the action proportional to : 
\begin{eqnarray}
\mathcal{I} \propto \int_\Sigma dr \, dt \,  \sum _{i=-2}^m l^{ i}\,   r^{-i} \,\left(  \alpha_i  \, b  \sqrt{1 + f^2} +\frac{\beta_i \, r}{ \sqrt{1 + f^2}} \, \left(a b'-  f \frac{\dot{a}}{a} + \frac{a b f f'}{1+f^2} \right) \right)  \,. 
\end{eqnarray}
Because the cross term $(2 b  f  \, dr \, dt)$ in the metric can be reabsorbed by a gauge transformation, we can derive the equations of motion with respect to $a$, $b$ and $f$, and then evaluate the result for $f=0$. Therefore, defining $\Delta_{\beta} =  \sum\limits_{i=-2}^m l^{ i} \,  r^{-i} \, \beta_i$ and $\Delta_{\alpha}  =\sum\limits_{i=-2}^m l^{ i} \,   r^{-i} \, \alpha_i $, the equations of motion of respectively $a$, $b$ and $f$ are :
\begin{eqnarray}
\begin{split}
\left(r  \Delta_\beta \right) b'= 0  \;, \;\; \Delta_\alpha -  \Big(   r  \Delta_\beta \, a \Big)'=0 \; , \;\;  -\left(  r \Delta_\beta \right) \frac{ \dot{a}}{a}  = 0 \, .
\end{split} 
\end{eqnarray} 
Comparing with the Weyl approach in Einstein gravity shows the close similarity between this first and our generalizations. 

\subsection{Vacuum solutions : Rational regular black holes}

The first equation of motion gives $b(t,r)=b(t)$ what can be reabsorbed by transformation of time coordinate, the third one gives $a(t,r)=a(r)$ and the second one :
\begin{eqnarray}
a(r)=\frac{-4 M +  \int \Delta_\alpha }{r \Delta_\beta} = 1 - \frac{4 M +  \left( r \Delta_\beta - \int \Delta_\alpha \right)}{r \Delta_\beta} \, ,
\end{eqnarray}
with $M$ the integration constant, and,
 \begin{eqnarray}
\int  \Delta_\alpha = \sum\limits_{i=2}^{m} r^{1-i} \, l^{\,i}  \left(\frac{ \alpha_{i}}{1-i}\right)  + l \,  \alpha_{1} \, \log\left(\frac{r}{l}\right)  + 2 \,r - \frac{2}{3} \, r^{3} \, \Lambda  \, .
\end{eqnarray} 
This proves that this model satisfies Birkhoff theorem (see \cite{DeserFrank}). Now, defining $\gamma_i=\beta_i - \frac{\alpha_i}{1-i}$, we get the general solution depending on the maximal order of correction, given by $m$, on the length scale of the Lagrangian $l$, on the integration constant $M$ and on the dimensionless sets of coupling constants $\alpha_i$ and $\beta_i$ : 
\begin{eqnarray}
\begin{split}
\label{NPGsol}
a = 1-\frac{ \frac{2}{3} \Lambda \, r^{m+2} +\Big( \left( \beta_{1} - \alpha_{1} \log\frac{r}{l} \right) l  + 4 M  \Big) r^{m-1} + \sum\limits_{i=2}^{m}  l^{ i}  \, r^{m-i}  \, \gamma_i }{2 \, r^m  + \sum\limits_{i=1}^m l^{ i}  \, \beta_i \, r^{m-i} } \, .
\end{split}
\end{eqnarray}
Except for the logarithmic term, these solutions are all the rational black holes with no mass terms at the denominator (unlike Hayward black hole for example) and they can be made regular if we consider corrections greater than the critical order $o=d$, as shown in \cite{RBHNPG}. In $4$ dimensions, we therefore need to consider at least an order $5$ correction, corresponding to $m=3$.

Note, en passant, that these models share some similarities with Lovelock-Lanczos gravity, because also here the critical order ($o=d$) scalar gives a topological invariant of the manifold. In our case, and for spherically symmetric spacetimes, if $\mathcal{S}_i$ is defined in (\ref{NPGscalarS}) without its normalization $\frac{1}{d-2-i}$, the term :
\begin{eqnarray}
\label{Topology}
\int_{\Sigma \times \Omega_r} d^{\, d}x \, \sqrt{-g} \, \mathcal{R} ^{(d-2)/2} \, \mathcal{S}_{d-2}  = \mathcal{A}(\Omega_{d-2,1})\, \int_\Sigma d^{2}x \, \sqrt{-\gamma} \, R(\gamma)  \,  ,
\end{eqnarray}
 gives (up to boundary terms) the Euler density of the $2$-dimensional manifold $\Sigma$ (and therefore its Euler characteristic), for any dimensions $d>3$, as shown in \cite{RBHNPG}, while for Lovelock-Lanczos gravity, the critical order scalar gives, for any background, the Euler density of the full $d$-dimensional manifold. This is the reason why we are using this normalization, so that the critical order scalar contributes to the equations of motion.

\subsubsection{First correction : Poisson-Israel regular black hole}

Now let's consider the smallest order of correction for which the black hole solution (\ref{NPGsol}) can be made regular, namely $m=3$ (odd order 5 correction). The corresponding solution is 
\begin{eqnarray}
a(r) = 1 - \frac{\frac{2}{3} \Lambda \, r^5 + r^2 \Big( 4 M + l \left( \beta_1 - \alpha_1 \log\frac{r}{l} \right) \Big) +\gamma_2 \, l^2 r  +\gamma_3 \, l^3 }{2 r^3 +\beta_1 \, l r^2  + \beta_2 \, l^2 r  +\beta_3 \, l^3 } \, .
\end{eqnarray}
However, as shown in \cite{RBHNPG}, we need to impose $\gamma_2=\gamma_3=\alpha_1=0$ and $\beta_3 \neq 0$ in order to have a (A)dS core at the center of the black hole, removing the singularity. Then, close to $r=0$, the solution looks like
\begin{eqnarray}
a(r\to 0)= 1 - \left(\frac{4 M + l \, \beta_1}{l^3 \, \beta_3}\right) r^2 + O(r^3)
\end{eqnarray}
Concerning the behavior of these cosmological regular black holes at infinity, the usual Schwarzschild-de Sitter asymptotic are slightly modified and it gives : 
\begin{eqnarray}
\begin{split}
a(r\to \infty) = &- \, \frac{\Lambda}{3}\, r^2 + \frac{1}{6} \Big( l \, \Lambda \, \beta_1 \Big) r + \Big( 1 - \frac{1}{12} l^2 \Lambda \left( \beta_1^{\,2} - 2 \beta_2 \right) \Big) \\
&+ \frac{-2M -\frac{\beta_1 \, l}{2} + \frac{l^3 \, \Lambda}{24} \Big( \beta_1^{\,3} - 4 \beta_1 \beta_2 + 4 \beta_3\Big)}{r} + O\left(\frac{1}{r^2}\right)
\end{split}
\end{eqnarray}
Therefore the UV corrections can have small, but non-vanishing effects in the IR. Indeed, we see that the modifications are combining both $l$ and $\Lambda$, i.e. the high and low energy scales describing these black holes, and are affecting both the Minkowskian part (of order unity) and the Schwarzschild part (of order $r^{-1}$), and add a term proportional to $r$. While these modifications could be interesting by allowing to constraint the value of the length scale $l$, they are nonetheless extremely weak, so we will only consider here a minimal modification to Schwarzschild-de Sitter behavior, setting then $\beta_2 =\beta_1=0$. What remains in (28) is only an effective mass term $M_{eff}$ :
\begin{eqnarray}
a(r\to \infty) = - \, \frac{\Lambda}{3}\, r^2 +1 - \frac{2 M_{eff}  }{r} + O\left(\frac{1}{r^2}\right)\, ,  \;\;\; \text{where} \;\; M_{eff} = M -\frac{l^3 \Lambda \, \beta_3}{12}  \, .
\end{eqnarray}

Let's set for now $\Lambda=0$. Moreover, as the only remaining dimensionless coupling constant is $\beta_3$, it can be reabsorbed into $l$, and for convenience we set $\beta_3=2$. The solution is therefore, 
\begin{eqnarray}
\label{PI}
a_{PI}(r) = 1 - \frac{2 \, M \, r^2}{r^3 + l^3} \; ,
\end{eqnarray}
which is the well-known Poisson-Israel regular black hole that has been found by a semi-classical argument involving the vacuum energy-density in \cite{PI}, and from a $2$-dimensional dilaton gravity model in \cite{TavesK} and \cite{Kunstatter}, where it was also studied in details. This shows that this non-singular spacetime can be found from the following (odd) order-$5$ correction to Einstein Hilbert action : 
\begin{eqnarray}
\mathcal{I}_{PI} =\frac{1 }{16\pi G }  \int_\mathcal{M} d^{4}x  \, \sqrt{-g} \left( R + 2 \,  \mathcal{V} \, \mathcal{R}^{3/2}\, \left( -2 \, \mathcal{R} + \mathcal{S}_3\right) \right) \, .
\end{eqnarray}

\subsubsection{Second correction : Non-minimal Einstein-Yang-Mills regular black hole}

Now let's turn to the second correction $m=4$. Now that we have seen the possible new behavior of these kind of regular black holes at infinity, we start with the following action, that consists only in an order-$(4+6)$ correction.
\begin{eqnarray}
\label{m=4}
\mathcal{I} =\frac{1 }{16\pi G }  \int_\mathcal{M} d^{4}x  \, \sqrt{-g} \left( R -2 \Lambda -\frac{1}{2} Q_m^{\,2} R(\Omega)^2 \left( 1 + q_b \left( 3 R(\Omega) -2 \mathcal{S}_4 \right) \right) \right)\, , 
\end{eqnarray}
where we have used $R(\Omega)=2\mathcal{R}$ and we have replaced the parameter $l$ by $q_b$ and $Q_m$ that have respectively the dimensions of an area and a length. The spherically symmetric solutions are : 
\begin{eqnarray}
\label{NMYM}
b(t,r)=b(t) \, , \;\;\; \text{and} \;\; a(r)=1-\frac{r^2 \left(\frac{\Lambda}{3} r^3 + 2 M r - Q_m^{\,2}\right)}{r^4 + 2 \, q_b \, Q_m^{\,2}} \, .
\end{eqnarray}
The replacement of constant is made in order to emphasized that this gravitational action has the same spherically symmetric gravitational equations of motion and solution that the so-called non-minimal Einstein-Yang-Mills theory (see \cite{BLZ} and \cite{NMYM2}), defined by the action
\begin{eqnarray}
\mathcal{I}_{NMYM} =\int d^4x \sqrt{-g} \Bigg( \frac{R - 2 \Lambda}{8 \pi} + \frac{1}{2} F_{ik}^{(a)}F^{ik(a)} + \frac{1}{2} F_{ik}^{(a)} F_{nm}^{(a)} \mathcal{R}^{ikmn} \Bigg) \, ,
\end{eqnarray}
where the gauge group is $SU(2)$, and $\mathcal{R}^{ikmn}$ is the non-minimal susceptibility tensor defined by :
 \begin{eqnarray}
\mathcal{R}^{ikmn}=q_b \Big( -\frac{1}{2} R \, g^{i [m}g^{n]k} + 2\left( R^{i[m}g^{n]k}+R^{k[n}g^{m]i} \right)- 6 R^{ikmn} \Big) \, .
\end{eqnarray}
The ansatz for solving the Yang-Mills field equations is a Wu-Yang monopole (see \cite{Zayats} for details). In this theory, the parameter $Q_m$ is the magnetic charge of this monopole (an integration constant), whereas in (\ref{m=4}), it is a dimensionful coupling constant. Imposing this monopole solution reduces the gravitational field equations to the same as those of (\ref{m=4}), meaning that its contribution to the spherically symmetric gravitational field equations can be reabsorbed into gravitational scalars, the first term $F_{ik}^{(a)}F^{ik(a)}$ corresponding to the order-$4$ curvature invariant $Q_m^{\,2} R(\Omega)^2$, what makes indeed the second term an order-$6$ one.

These regular black holes were studied in details in \cite{BLZ} and \cite{NMYM2}. An interesting feature that was noted is that, since $q_b>0$ in order to have regular solutions, their behavior close to the origin is : 
\begin{eqnarray}
a(r\to 0) = 1 + \frac{r^2}{2 q_b} + O(r^3) \, .
\end{eqnarray} 
We see that these black holes have an AdS core, instead of the usual dS one, what produces a small cavity close to $r=0$. Note also that they have the standard Schwarzschild-dS asymptotic :  
\begin{eqnarray}
a(r\to \infty) = - \, \frac{\Lambda}{3}\, r^2 +1 -\frac{2M}{r} + O\left(\frac{1}{r^2}\right) \, . 
\end{eqnarray} 

\subsubsection{Limiting curvature conjecture \& physical regular black holes}

Up to now, we have tried to impose some simple requirements at the level of the action, in order to see what kind of theory and solution could come out. We have found that a minimal generalization of EH action in spherical symmetry, in the form of an effective-like action, admits as solutions all the rational regular black holes with no mass term at the denominator.

This last point is crucial because it implies that all these RBHs have a well-known pathology (see for example \cite{Frolov_3}) : they do not have bounded invariants with respect to the mass $M$. Indeed, the Ricci scalars of the PI and NMYM solutions behave respectively as follows : 
\begin{eqnarray}
R_{PI}= \frac{12 l^3 M \left( 2l^3 - r^3\right)}{\left(l^3 + r^3 \right)^3}\underset{r \to 0}{\longrightarrow} \frac{24 M}{l^3} \underset{M \to \infty}{\longrightarrow} \infty
\end{eqnarray}
and setting to zero the cosmological constant,
\begin{eqnarray}
R_{NMYM}= \frac{8 q_b \, Q_m^{\,2} \left(-6 q_b \, Q_m^{\,4} + 5 Q_m^{\,2} r^4 + M \left(20 q_b \, Q_m^{\,2} r - 6 r^5 \right) \right)}{\left(2 q_b  Q_m^{\,2} + r^4\right)^3}\underset{r \to 0}{\longrightarrow} -\frac{6}{q_b} 
\end{eqnarray}
Here a distinction can be made : the PI solution has not even a bounded curvature at $r=0$, while the NMYM one does. However, for non vanishing finite radius $r$, both behave as : $R\underset{M \to \infty}{\longrightarrow} \infty$, and it is the same for the other invariants, and for higher order $m$ solutions of this model. This unboundedness makes the inner horizon of these black holes arbitrary small as the mass grows, as one can checked by solving $a(r_{\text{inner}})=0$. If we consider, for simplicity, that $l$ is small compared to the mass $M$, then it gives for the PI solution (\ref{PI}) : $r_{\text{inner}} \approx l  \sqrt{\frac{l}{2 M}} + O\left( \frac{l^3}{ M^2}\right) \underset{M \to \infty}{\longrightarrow} 0$, and $r_{\text{inner}} \approx l  \left(\frac{l}{2 M}\right)^{1/3} + O\left( \frac{l^2}{ M}\right) \underset{M \to \infty}{\longrightarrow} 0$ for the NMYM one (\ref{NMYM}), where we set $q_b=l^2$ and $Q_m=l$ for simplicity. Therefore there is no minimal size for the (A)dS cores of these black holes, and the singularity is recovered as $M\to \infty$. 
\\
\\
More generally, we could say that regular black holes can be divided into two classes, one for which the curvature is unbounded with respect to the mass (as in our solutions, among which the Poisson-Israel and Non-minimal-Yang-Mills RBHs, and also for example the non-commutative geometry inspired models \cite{NSS}), and another class for which the curvature is bounded. For example the well-known Hayward regular black hole \cite{Hayward_1} is precisely designed in order to cure the pathology of the Poisson-Israel one. The metric and its associated Ricci scalar are :
\begin{eqnarray}
\label{Hayward}
a_{H}(r)= 1 - \frac{2 \, M \, r^2}{r^3 + 2 \, M \, l^2} \, , \; \text{and} \;\; R_H = \frac{24 l^2 M^2 \left(4l^2 M - r^3\right)}{\left(2l^2 M + r^3 \right)^3} \underset{M \to \infty}{\longrightarrow} \frac{12}{l^2} \, ,
\end{eqnarray}
which satisfies the Limiting Curvature Conjecture, meaning that all the invariants are bounded by one universal constant $l$ (see \cite{M,Mar}
 and \cite{Frolov_6}), which in turn, makes the inner horizon radius bounded from below by this constant $l$. Indeed, when $l$ is small compared to the mass $M$, it gives : $r_{\text{inner}} \approx l + O\left(\frac{l^2}{ M} \right)$.
\\

From this observation, there are three possible directions to go : either one wants to consider unbounded RBHs, and thus has to reject the LCC, or one finds a mechanism that provides a natural cut-off mass $M_{max}$ which would allow to have bounded curvature invariants for the first class, or one considers only the second class, in which the Hayward spacetime belongs, but then one needs to understand what types of action produce these solutions. In \cite{RBHNPG}, we proposed a possible argument involving the presence of a positive cosmological constant, $\Lambda >0$, in order to have such maximal mass $M_{max}$ and thus, to have bounded invariants for the first class of regular black holes.
\\
\\
However, the third possibility, i.e. to directly consider bounded RBHs like the Hayward one, is the simpler one and also the more satisfactory. Because, as we saw, RBHs that do not satisfy the Limiting Curvature Conjecture (bounded curvature invariants by a single constant), do not satisfy neither the Bronstein argument (see \cite{BRON1,BRON2} and \cite{ROV} for a modern formulation), stating that there should be theoretically a minimal measurable size for BHs (and therefore, following the standard view about quantum mechanics, a physical minimal size, below which length does not make sense anymore). This argument is at the core of quantum gravity, as it combines both quantum field theory and general relativity concepts. However, for unbounded RBHs, we saw that the location of the inner horizon can be arbitrarily smaller than the Planck length, and therefore they do not encode this fundamental semi-classical argument. Fortunately, the non-polynomial gravity approach can also provide a general framework to find this type of regular solutions as we will see soon.
\\
\\
Before doing so, note that such a general framework is not only needed to find bounded regular black holes. In fact, there are many criterions in the literature that have been studied in order to understand what could be a self-consistent regular black holes. To list a few, such a physical RBH should solve the mass inflation problem of inner horizons, see for example \cite{MII,Poi,LQGMI,Frolov_2}, where it was found in this last that to do so, it might require a non trivial redshift function (corresponding in gauge (\ref{GAUGE}), to $b(t,r)$), encoding the time delay between the center and infinity. There is also other Cauchy horizon instabilities that should be cured \cite{Mae}. Another criterion, would be for the effective metric to describe a Planck star (see \cite{PLANCK}, what requires also a non-trivial redshift function \cite{LOR1,LOR2}. In this paper, it was also noted that a physical regular black hole should behave at infinity in accordance with the quantum corrections to the Newtonian potential. Of course, the study of energy conditions is also an important ingredient to study the physicality of RBHs. Finally, note that there also exist alternatives to static regular black holes, for example effective metric fields that describe a bounce under gravitational collapse of matter (see 	\cite{Stellar,Bambi1,Bambi2,Bambi3,Bambi4,Casadio}). In this last paper, it was suggested that bounces and static RBHs could both be different final states of a gravitational collapse (if static RBHs are possible at all), depending on the initial conditions of the collapsing system.

\subsection{A Unified Approach}

In order to have a general tool to build black hole solutions corresponding to the previous requirements, we summarize here a proposition of unified approach (again, see \cite{RBHNPG} for details), based on a work by Kunstatter, Maeda and Taves \cite{Maeda}.

In this paper, the authors have generalized the $2$-dimensional spherically symmetric reduction of the Lovelock-Lanczos action as the following scalar-tensor theory that leads to second order equations of motion : 
\begin{eqnarray}
\label{KMT2}
\mathcal{I}_{KMT}^{(2)}=\frac{\mathcal{A}_{(d-2)}}{16 \pi G} \int_\Sigma d^{2}x \, \sqrt{-\gamma}  \, \left( \phi\left(r\right) R\left(\gamma\right) + \eta\left(r,Z\right) + \chi\left(r,Z\right) \frac{\gamma^{AB} \tilde{\nabla}_A r \tilde{\nabla}_B Z }{Z} \right),
\end{eqnarray}
where $Z = \gamma^{AB} \tilde{\nabla}_A r \tilde{\nabla}_B r$. 
The question was raised to understand what subsets of these models can be lifted to higher dimensions, that is, what $d$-dimensional theory would give the action (\ref{KMT2}) after restricting the action to spherical symmetry and integrating out the angular part of the action to give the factor $\mathcal{A}_{(d-2)}$. Quite surprisingly, if we consider non-polynomial curvature invariants, the answer to this question is that the whole action can be lifted to any $d>3$ dimension by considering the following action (see \cite{RBHNPG} for the general $d$-dimensional case and the explicit construction): 
\begin{eqnarray}
\label{KMT4}
\mathcal{I}_{KMT}^{(4)} = \frac{1 }{16\pi G } \int_\mathcal{M} d^{4}x \, \sqrt{-g} \Bigg( \Phi\left(\mathcal{R}\right) R\left(\Sigma\right) + \mathcal{N}\left(\mathcal{R},\mathcal{Z}\right) + \mathcal{X}\left(\mathcal{R},\mathcal{Z}\right) \frac{\nabla^\alpha \omega_{\alpha\beta} \nabla^\beta \mathcal{Z} }{\mathcal{Z}} \Bigg), 
\end{eqnarray}
where $\mathcal{R} \bigr\rfloor = 1/r^2$ is defined by Eq(\ref{NPGscalarR}) in terms of $u_{\alpha\beta}$, and $\omega_{\alpha\beta}$ is the $4$-dimensional degenerate metric of $\Sigma$, built from $u_{\alpha\beta}$ as follows :
\begin{eqnarray}
\omega_{\alpha\beta}  :=  - \left( \frac{g_{\alpha\beta} + 4 \, \,  u_{\alpha\beta} }{3} \right) \, .
\end{eqnarray} 
From it, one can build the following scalars : 
\begin{eqnarray}
 \mathcal{Z}= \nabla_\gamma \omega_{\alpha\beta} \nabla^\gamma \omega^{\alpha\beta}   \, ,  \;\; \text{with} \;\;   \mathcal{Z} \Bigr\rfloor =  4 \left( \frac{Z }{r^2}  \right),
  \end{eqnarray}
which is just a kinetic term in which, roughly speaking, $\omega_{\alpha\beta}$ plays the $4$-dimensional role of the $2$-dimensional scalar field $r(x)$,
\begin{eqnarray}
R(\Sigma) &:= \left(R^{\alpha\beta}+\nabla^\alpha \nabla^\beta \right) \omega_{\alpha\beta} - \frac{\mathcal{Z}}{2} \, ,  \;\; \text{with} \;\;   R(\Sigma) \bigr\rfloor = R(\gamma)    \, ,
\end{eqnarray}
which plays the $4$-dimensional role of the Ricci scalar of the $2$-dimensional manifold $\Sigma$, and finally,
\begin{eqnarray}
\left( \nabla_\alpha \omega^{\alpha\beta} \nabla_\beta \mathcal{Z} + \mathcal{Z}^2 \right) \Bigr\rfloor =8 \left( \frac{\gamma^{AB} \tilde{\nabla}_A r \tilde{\nabla}_B Z }{r^3}\right)\,.
\end{eqnarray}
From these definitions, it is clear that the spherically symmetric sector of the action (\ref{KMT4}) is the same as  action (\ref{KMT2}).
\\
\\
Moreover, from this last, many bounded regular black hole solutions were found in \cite{Maeda}, from the Hayward one Eq(\ref{Hayward}), to a new one : 
\begin{eqnarray}
a(r) = 1 + \frac{r^5}{ l^4 M }\left(1- \sqrt{1+ \frac{4 M^2 l^4}{r^6}} \right)  .
\end{eqnarray}
Interestingly, still in the mentioned paper by K.M.T., these two were also given an effective-like action interpretation, called Designer Lovelock, as a summation, up to infinite order, of the Lovelock-Lanczos action, for suitably chosen coupling constants. Therefore, there exists a particular $4$-dimensional action (\ref{KMT4}), involving non-polynomial curvature scalars, that has also these bounded regular black hole solutions and an effective-like action interpretation (a generalization of Lovelock-Lovelock gravity) in spherically symmetric spacetimes.

The drawback being that one needs necessarily to reconstruct the action from given solutions, or to fix an infinite number of coupling constants, in order to find (regular) solutions. Despite its own problems, it was not the case of the minimal generalization of Einstein theory presented before. 

\subsection{Cosmological bounce}

We have just seen that by extending Lovelock-Lanczos gravity, in any dimensions, to higher orders than the critical order $o=d$ (above which all the Lovelock scalars vanish identically), it is possible to get regular solutions. This is only possible by considering non-polynomial scalars that allow to circumvent Lovelock theorem \cite{Lovelock_1,Lovelock_2,Lovelock_3}.

It was found in \cite{Helling_3,Date} that this is also the case for cosmology : a Lovelock-like modification to the Friedmann equations (i.e. a polynomial correction in the Hubble parameter $H= \frac{\dot{a}}{a}$, see \cite{cosmoLove1,cosmoLove2,cosmoLove3}), that is extended beyond the critical order, is able to reproduce the loop quantum cosmology bounce solution, and therefore to solve the singularity problem in the cosmological context. Just like before, the drawback is that an infinite number of coupling constants have to be fixed. In \cite{CCZ17}, we showed that these Lovelock-like modifications admit a very simple non-polynomial gravity formulation as we will see now. 
\\
\\
First, because the Cotton tensor is vanishing identically for homogeneous and isotropic metric fields, defined by the interval
\begin{eqnarray}
\label{FLRW}
 ds^2 = - N(t) dt^2 + a(t)^2 d\vec{x}^{\,2} \, , 
\end{eqnarray}
we need to consider another order-$0$ tensor. As shown in \cite{CCZ17}, we can use the following one (that is also an order-$0$ tensor for static spherically symmetric spacetimes): 
\begin{eqnarray}
\label{Order0}
  V_\alpha := \frac{\partial_{\alpha} R}{\sqrt{-\partial_{\sigma} R\partial^{\sigma} R} }   \;\;\;\;\; \text{and} \;\;\;\;\;   V_{\alpha\beta}  :=  V_{\alpha} V_{\beta}  \, .
\end{eqnarray}
from which we can construct second order scalars :
\begin{eqnarray}
\label{Scalars}
 K  := \frac{1}{9} \big( \nabla^{\alpha}\nabla^{\beta} V_{\alpha\beta} - V_{\alpha} \nabla^{\alpha} \nabla^{\beta} V_{\beta} \big) \;\;\;\;\; \text{and} \;\;\;\;\;  \Omega  := \frac{R}{6}-2 K \,,
\end{eqnarray}
such that
\begin{eqnarray}
  K \Bigr\rfloor =\frac{H^2}{N}  \;\;\;\;\; \text{and} \;\;\;\;\;  \Omega \Bigr\rfloor  =\frac{\dot{H}}{N} -  \frac{H \,
    \dot{N}}{2 N^2} \,.
\end{eqnarray}
Then we consider the following action :
\begin{eqnarray}
 \mathcal{I} =\frac{1}{16\pi G}\int d^4x \sqrt{-g} \left(R -2 \Lambda +  \sum_{i=1}^{\infty } (-1)^{i+1}  \binom{1/2}{1+i} \; S^i \; \Omega \right) + \mathcal{I}_m,
\end{eqnarray} 
where $\mathcal{I}_m$ is the matter action, $\binom{n}{m}$ is the generalized binomial coefficient defined by $\binom{n}{m} :=\frac{\Gamma (n+1)}{\Gamma  (m+1) \Gamma  (n-m+1)}$, and we have defined the dimensionless curvature invariant $S=\frac{3}{2 \pi \rho_c}\, K$, with $\rho_c$ playing the role of critical density.

Deriving the equation of motion with respect to $N(t)$ and then setting $N(t)=1$ due to time reparametrization invariance, leads to the following modification of Friedmann equation : 
\begin{eqnarray}
\label{EOM}
4 \pi  \rho_c \left(1-\sqrt{1-\frac{3 H^2 }{2 \pi  \rho_c}}\right) =8 \pi   \rho+\Lambda\, .
   \end{eqnarray}
Expending in $\rho_c$, it is clear that it is just a Lovelock-like polynomial correction to Friedmann equation, that goes up to infinite order, and with suitably chosen coupling constants. In this sense, this is the analogue of the Lovelock Designer theories found in \cite{Maeda}.

Now, solving in $H^2$ gives the well-known loop quantum cosmology modification \cite{Ashtekar,Ashtekar_2,bo,Bojo}
\begin{eqnarray}
\label{MF}  
H^2=\, \frac{8 \pi \bar{\rho}  }{3} \Big( 1 -  \, \frac{ \bar{\rho}  }{\rho_c} \Big),
\end{eqnarray}
where $\bar{\rho} = \frac{ \Lambda}{8\pi} + \rho $. It was shown in \cite{CCZ17} that, considering a perfect fluid with equation of state $p=w \rho$, the solutions to this equation and the energy conservation equation
 \begin{eqnarray}
 \label{Cons}
\frac{d\rho}{dt}+3H \big(\rho + p \big)=0\,,
\end{eqnarray}
are
\begin{eqnarray}
\label{SOLUTION}
a(t) =a_0 \left(\frac{-1+2
   \Lambda  \mu +\cosh \Big(\left(\sqrt{3}  (1+w) t
   \right) \sqrt{\Lambda } \sqrt{1-\Lambda  \mu
   }\Big)}{2 \Lambda  (1-\Lambda  \mu )}\right)^{\frac{1}{3 (1+w)}},
\end{eqnarray}
and 
\begin{eqnarray}
\tilde{\rho}(t)=-\frac{2 \Lambda  (-1+\Lambda  \mu )}{-1+2 \Lambda  \mu
   +\cosh \left(\left(\pm \sqrt{3} t (1+w)  \right)
   \sqrt{\Lambda } \sqrt{1-\Lambda  \mu }\right)},
\end{eqnarray}
where $ \tilde{\rho}=8 \pi  \rho$, $\mu=\frac{1}{8 \pi  \rho_c}$ and $a_0$ is an integration constant. These are regular solutions, providing that $\mu \neq 0$.
\\

Note that, in the context of modified gravity models, this correction to the Friedmann equation and the associated bounce solution has also been found in the context of mimetic scalar-tensor theories in 	\cite{H,Mukhanov_2,Mukhanov} (see \cite{mi_1,Muk0,mi_6,mi_2,mi_3,mi_4,mi_5,mi_7} for more details about mimetic gravity, while its generalization to $F(R)$ and for $F(R)$ ghost-free models see \cite{N3,N4}.  Furthermore, polynomial and second order corrections to Friedmann equation have been found from a Galileon inspired action in \cite{Gali} and from non-polynomial invariants in \cite{gao}.

\section{ The fluid approach}

Finally, we will conclude this review by a popular approach to deal with the singularity issue : to reconstruct the stress-energy tensor, i.e. the matter content, possibly exotic, given regular ansatz. We present the approach, with some additional remarks on the method. It is also worth to notice that the fluid approach can be viewed as an effective version of a Lagrangian approach, where the whole covariant form of the Lagrangian is still inaccessible.

Let us  introduce the effective density $T^0_0=-\rho$, the effective radial pressure $p_r=T_1^1$ and the effective tangential pressure $p_T=T^2_2=T^3_3$, $T_{\mu\nu}$ being the components of the stress-energy tensor. As a result, in the gauge (\ref{metric0}), the  gravitational equations reduce to
\begin{equation}
\label{ege1}
rg'+g-1 = - 8\pi r^2 \rho\,,
\end{equation}
\begin{equation}
\label{ege2}
rf'+f-\frac{f}{g} = 8\pi\frac{f}{g} r^2 p_r\,,
\end{equation}
\begin{equation}
\label{ege3}
rf''\frac{g}{f}+g(\frac{f'}{f}+\frac{g'}{g})+\frac{rg}{2g}(\frac{f'g'}{fg}-\frac{f'^2}{f^2}) = 16 \pi r p_T\,.
\end{equation}
These equations must be supplemented by an equation of state, say $p_r=\omega \rho$.

It should be noted that if one is looking for solutions with $f=g$, then there is a drastic simplification and, as consequence, one has $p_r = - \rho $ and $ p_T=-\rho-\frac{r}{2}\rho'$, this being equivalent to stress tensor conservation $\nabla^{\mu} T_{\mu \nu}=0$ (Bianchi identity).

As a result, the general solution reads
\begin{equation}
\label{ege8}
f(r) = f_1 (r) - \frac{C}{r} =1-\frac{C}{r}-\frac{8\pi}{r}\int_0^r dr_1 r_1^2\rho(r_1)= 1-\frac{C}{r}-\frac{2m(r)}{r} \,.
\end{equation}
The quantity $m(r)$ is called mass function. 

If the effective density $\rho$ does not depend on $f(r)$, and its derivatives, it cannot depend on the constant $C$, and as a consequence the model described by the above gravitational equations contains the $\frac{C}{r}$ term, and this leads to a central singularity in $r=0$, as soon as $C$ is not vanishing. This is  independent on the  contribution coming from $f_1(r)$. This result is known within the NED approach (see for example \cite{breton} and original references cited therein).   

Furthermore, after having chosen $C=0$, invoking a suitable boundary condition, one has to make a suitable choice for  $\rho(r)$ in order to have a de Sitter core at $r=0$.  Well known examples are the choices $\rho(r)\equiv \rho_0 e^{-\frac{r^3}{a^3}}$, and $\rho(r) \equiv\frac{A}{(4\pi \theta)^{3/2}} e^{-\frac{r^2}{4 \theta}}$, which correspond to the Dynmikova \cite{Dymnikova92} and Spallucci et al. \cite{NSS} regular black holes.

Some remarks are in order. First, within this fluid approach, one should observe that the radial speed of sound reads $v_r^2=\frac{d p_r}{d\rho}=-1$, thus is negative. As a consequence, in general, the static regular solution is unstable \cite{Bhar:2017ynp}.

Since $f=g$ only if $p_r = -\rho$ and this generates an imaginary speed of sound, one should drop the $f=g$ condition; however, in this case the Einstein equations are much more difficult to solve and this general case will not be considered here.

\section{ Discussion}

The results reviewed in this paper can be divided into two categories. The first one, studied in Section 1 and 3, are about modifications of the matter content of General Relativity, introduced in such a way that the singularity theorem assumptions are violated. While in section 2, we saw some geometrical modifications of Einstein Equations. One of the advantage of modifying the Geometrical content of the theory is that, by doing so, regular solutions are found already when a point-like mass is considered. Therefore, given more regular distributions of matter, these models might still provide regular solutions. On the contrary, if one modifies directly the matter content in a given way, it remains to be explained why every kind of matter should behave in this precise way at small distances. In this respect, non commutative geometry inspired models \cite{NSS} are an exception, as they provide such an explanation.

More precisely, we saw in Section 1 that NED is about reconstruction of non-linear electrodynamics Lagrangians, given regular black holes ansatz. To our knowledge, there is no arguments to impose on the Lagrangian, in order to \textit{find} regular solutions within this approach. Nonetheless, it is a powerful tool to reconstruct both regular solutions and an associated Lagrangian description in terms of a fundamental field. Moreover, even rotating solutions have been reconstructed from this approach \cite{ROTATING}, which is a key step toward a better understanding of physical RBHs.

Similarly, we saw in Sec. 3 that many regular black holes found in the literature can be viewed as ansatz, from which some effective fluids are reconstructed to get the given solutions. Moreover, when $f=g$ in (\ref{metric0}), all the solutions present an instability due to a negative sound speed. 
\\

In Section 2, we summarized the non-polynomial approach \cite{RBHNPG}, and we have seen that it can constitute a unified approach to study $d$-dimensional effective spherically symmetric metric fields, in particular static RBHs, coming from a modification of the gravitational sector of Einstein field equations. In this way, RBHs can also be studied and classified in terms of the actions from which they are derived. In some cases, when the action is polynomial in spherical symmetry, it is also possible to associate an order to the corrections of Einstein Equations : for example we have seen that the family of Non-minimal Yang-Mills RBHs correspond to order $4+6$ gravitational corrections, while the Poisson-Israel one corresponds to an order $5$ correction. These regular black hole solutions are derived from a minimal generalization of Einstein gravity in spherically symmetric spacetimes, where the curvature invariants that modifies it can be interpreted as powers of the Ricci scalar of the horizon manifold, multiplied by a suitable scalar able to preserve the GR structure in spherical symmetry, and related to the Euler characteristic of the 2D submanifold.
\\

However, some pathologies are present in these RBHs, namely, they do not satisfy the Limiting Curvature Conjecture, and thus, they have unbounded curvature with respect to the mass, and the location of their inner horizons is not bounded from below. This is a very general issue about static regular black holes, and many others found in the literature suffer from the same pathology.

We reviewed an NPG approach that allows to find a four dimensional expression of two dimensional dilaton gravity reduction models, used to study the quantum properties of black holes such as their evaporations. Because these models admit bounded regular black holes, so does the NPG generalization. In particular, the Hayward metric and recently proposed regular black hole are bounded, and are possible to derive from a generalization of Lovelock-Lanczos gravity, beyond the critical order. This is also the case, in the context of cosmology, of the LQC bounce solution. Indeed, we presented a NPG modification to the Friedmann equation, that also involve Lovelock-like terms, and upon choosing suitably an infinite set of dimensionless coupling constants, it is able to mimic the LQC bounce solution.

From these two sectors, homogeneous and isotropic cosmology and spherical black holes, we see that Lovelock-type modifications to Einstein field equations, providing that it goes beyond the critical order, where normally all the Lovelock scalar are identically vanishing, is a quite efficient way to produce regular solutions.
\\

Within these 2D models and their NPG generalization, and within NED and fluid approaches, it remains to be seen if bounded RBHs can be found to solve the mass-inflation problem and other instabilities, inherent to the presence of an inner horizon. Moreover, even if possible, one needs to understand also if a formation of these static RBHs by gravitational collapse is possible, and if rotating solutions exist.

It should also be observed that there exist other modified gravity models like $F(R)$, $F(G)$, non-minimal $F(R)$-matter models (for a review see \cite{N1,N2}) in which $G$ is the Gauss-Bonnet invariant, and able to describe inflation, bounce and late acceleration. For these models, in principle, it should be possible to find regular solutions. In fact, making use of the results contained in \cite{gao}, it is not difficult to write down an action, which within a flat FLRW space-time leads to modified Friedmann equation (\ref{MF}). However, this action  is only fine  for the flat FLRW space-times, since in the case of a static black hole can produce field equations of order higher than 2. On the other hand, our NPG Lagrangians work well also in spherically symmetric static space-times.

It should also be noted that the whole discussion within the framework of an $F(R)$ or $F(G)$ theory is hugely complicated and turns out to be difficult to obtain fully analytical results (see e.g. \cite{kanti}, \cite{kanti2}).

Finally, we presented in the Appendix a covariant form of the Sakharov criterion to probe if spacetimes are singular or not. It would also be interesting to try to extend this result to the rotating case, what might also help to understand what kind of theories would admit rotating regular black holes.

\section{Appendix}

We start this Appendix with a review of some standard topics, necessary to discuss a covariant form of the regularity criterion, which reduces to the standard Sakharov criterion for static spherically symmetric spacetimes.

\subsection{Spherically symmetric spacetimes}

We mainly restrict our analysis on spherically symmetric dynamical spacetime, admitting a dynamical horizon. In the following, for the sake of completeness,  we briefly review the general formalism we shall be dealing with \cite{kodama,sean09,bob09,noi11}. Recall that any spherically symmetric metric can locally be expressed in the form (here in four dimensions)
\beq
\label{metric2}
ds^2 =\gamma_{AB}(x)dx^Adx^B+ r^2(x) d\Omega_k^2\,,\qquad A,B \in \{0,1\}\;,
\eeq
where the two-dimensional metric
\beq d\Sigma^2=\gamma_{AB}(x)dx^Adx^B
\label{nm}
\eeq
is sometimes referred to as the normal metric, $\{x^A\}$ are associated coordinates and $r(x)$ is the areal radius, considered as a scalar field in the two-dimensional normal space. As for $d\Omega_k^2$, it is a maximally symmetric space, whose metric is
\beq
\label{hmetric}
d\Omega_k^2=\frac{d\rho}{1-k\rho^2}+\rho^2 d\phi^2 \,,\qquad i,j \in \{0,1\}\;.
\eeq
Thus, for $k=1$, one has the two sphere $S^2$, for $k=0$ the two-dimensional torus $T^2$, and for $k=-1$, the two dimensional Riemann Surface $H^2/\Gamma$.

A relevant scalar quantity in the reduced normal space is 
\beq
Z(x)=\gamma^{AB}(x)\partial_A r(x)\partial_B r(x)\,, \label{sh} 
\eeq 
since the dynamical trapping horizon, if it exists, is defined by
\beq 
Z(x)\Big\vert_H = 0\,, \label{ho} 
\eeq
providing that $\partial_A Z \vert_H \neq 0$.

Another important normal-space scalar is the Hayward surface gravity associated with this dynamical horizon, given by 
\beq
\kappa_H=\frac{1}{2}\tilde{\Box} r \Big\vert_H\,. \label{H} 
\label{H}
\eeq 
This is a generalization of the usual Killing surface gravity. In a spherically symmetric dynamical case, it also is possible
to introduce the Kodama vector field $K$. Given the metric (\ref{metric2}), the Kodama vector components are
\beq 
 K^A(x)=\frac{1}{ \sqrt{-\gamma}}\,\varepsilon^{AB}\partial_B r\,,
\qquad  K^\theta=0= K^\varphi \label{ko} \,, 
\eeq 
where $\varepsilon^{AB}$ is the usual fully antisymmetric tensor. We may introduce the Kodama trajectories, and related Kodama observer, by means of integral lines of Kodama vector
\beq
\frac{d\, x^A}{d \lambda}= K^A= \frac{1}{ \sqrt{-\gamma}}\,\varepsilon^{AB}\partial_B r\,. 
\label{ko1}
\eeq
As a result, $\frac{d\, r(x(\lambda))}{d \lambda}=0\,$. Thus, in a  generic spherically symmetric spacetimes, the areal radius $r$ is conserved along Kodama trajectories. In other word, a Kodama observer is characterized by the condition $r=r_0$. The operational interpretation goes as follows. Static observers in static BH become in the dynamical case Kodama observers whose velocity 
\beq
v^A_K=\frac{K^A}{\sqrt{Z}}\,, \quad \text{such that} \quad \gamma_{AB}v^A_Kv^B_K=-1 \,.
\eeq
The energy measured by this Kodama observer at fixed areal radius $r_0$ is
\beq
E=-v^A_K\partial_A I=-\frac{K^A \partial_A I}{\sqrt{Z_0}}=\frac{\omega}{\sqrt{Z_0}} \,,
\eeq
where $I$ is the classical action of the relativistic particle and $\omega=-K^A\partial_AI$, its Kodama energy, and  $\partial_AI$ being its momentum.

\subsection{Singular and regular spacetimes }

As mentioned in \cite{Seno},  if the spacetime is sufficiently smooth, and  one assume suitable geometric condition on the Ricci tensor, plus causality and boundary conditions then there exist non spacelike inextensible geodesics. This is usually accepted as the presence of singular spacetimes. For a complete discussion see the well known monography \cite{Hawking_Ellis} and \cite{Seno}. For the inflationary case see also \cite{Li} and the papers quoted there.  

\subsubsection{Non-spacelike geodesics in curved spacetimes}

Now, we recall  the well known derivation of geodesics equation related to massless (light-like) geodesics and massive (time-like geodesics). To begin with, let us start from the generic metric
\begin{equation}
ds^2=g_{\mu \nu}(x)dx^{\mu}dx^{\nu} \,.
\end{equation}
Let us denote by $\dot{ x^\mu}=\frac{d x^\mu}{d \lambda}$, $\lambda$ an affine parameter. The geodesic equation for non-spacelike geodesics may be derived from the Lagrangian
\begin{equation}
L=- \frac{g_{\mu \nu}(x)\dot x^{\mu} \dot x^{\nu}}{2V}+\frac{m^2}{2}V \,.
\end{equation}
where $V$ is the einbein, a Lagrangian multipliers,  implementing the reparametrization invariant and $m^2$ is a mass like term, which is positive for timelike geodesics and vanishing for lightlike geodesics. In fact, the variation with respect to $V$ gives
\begin{equation}
g_{\mu \nu}(x)\dot x^{\mu} \dot x^{\nu}=-m^2V^2 \,.
\end{equation}
Now for lightlike geodesics, we have $m^2=0$, while for timelike geodesics we may take $V^2m^2=-1 $.
The geodesics equations can be obtained making the variation with respect to $x^\mu$, namely
\begin{equation}
  \frac{ d}{d \lambda} \left(g_{\mu \alpha}(x)\dot x^{\alpha}\right) =\frac{1}{2}\partial_{\mu} \left(g_{ \alpha \beta}(x)\dot x^{\alpha}\dot x^{\beta}  \right) \,.
\end{equation}

\subsubsection{Static Spherically Symmetric spacetimes}

After these general discussion, here we deal with some examples of SSS spacetimes. For an SSS spacetimes in the Schwarzschild diagonal gauge, namely 
\begin{equation}\label{sss}
 ds^2 = -b^2(r) f(r) dt^2  + \frac{ dr^2}{f(r)} + r^2 dS^2\,.
\end{equation}
a well known criterion is the existence of a de Sitter core at the center, namely for small $r$, one should have $f(r)= 1-Ar^2+...$.

Another criterion has been presented in \cite{Seno}. In  diagonal gauge,  
\begin{equation}\label{sssS}
 ds^2 = -F(\rho)dt^2  +  d\rho^2 + r^2(\rho) dS^2\,,
\end{equation}
for small $\rho$, one has to have $r(\rho)= \rho-A\rho^3+...$, and $F(\rho)$ smooth, namely
$F(\rho)= F_0+O(\rho)$.

\subsubsection{The FLRW spacetime}

Now let us recall the FLRW spacetime form
\begin{equation}\label{frw}
 ds^2 = -dt^2 + a^2(t) \left( \frac{dr^2}{1-R_0r^2} + r^2 d\Omega^2 \right)\,.
\end{equation}
Then, the geodesic equation reads
\begin{equation}\label{frw1}
\frac{- d \dot t}{d \lambda}=H(t)\left((\dot t)^2-\varepsilon \right)\,,
\end{equation}
where for  timelike trajectories $\varepsilon=1$, and for lightlike ones $\varepsilon=0$.
The solution of  above differential equation reads
\begin{equation}\label{frw2}
\dot t=\frac{d t}{d \lambda}= \frac{\sqrt{a_0^2+\varepsilon a(t)^2}}{a(t)}\,.
\end{equation}

\subsubsection{Past-incompleteness in FLRW}

The above result permits a quite simple discussion on the geodesic completeness in FLRW spacetimes. In fact, we may rewrite
\begin{equation}\label{frw2}
\frac{d \lambda}{d t}=\frac{a(t)}{\sqrt{a_0^2+\varepsilon a(t)^2}}\,.
\end{equation}
If $a(t)$ is always positive and never vanishing for all $t$, it follows that $\lambda$ is a monotone function, and one has the geodesic completeness. Furthermore, a further integration gives
\begin{equation}\label{frw3}
 \lambda=C_0+\int dt \frac{a(t)}{\sqrt{a_0^2+\varepsilon a(t)^2}}\,.
\end{equation}
If $a(t)>0$, we may write
\begin{equation}\label{frw4}
 \lambda=C+\int_{-\infty}^t dt' \frac{a(t')}{\sqrt{a_0^2+\varepsilon a(t')^2}}\,.
\end{equation}
If this integral is {\it divergent}  all non space-like geodesics are past-complete, and no singularities are present.

If $a(t)$ is vanishing in the past, say $a(0)=0$, we may write
 \begin{equation}\label{frw5}
 \lambda=C+\int_{0}^t dt' \frac{a(t')}{\sqrt{a_0^2+\varepsilon a(t')^2}}\,.
\end{equation}
If this integral is {\it convergent}, one has past incomplete geodesics, and thus singularities are present.

 First elementary example is the flat FLRW bounce solution
\begin{equation}\label{frw6}
a(t)=1+A^2t^2 >0 \,.
\end{equation}
In this case $\varepsilon=0$, and the integral (\ref{frw4}) is divergent, thus no singularities.

The second one is 
\begin{equation}\label{frw7}
a(t)=t^{\alpha}  \,, \quad \alpha>0\,.
\end{equation}
In this case,  the integral (\ref{frw5}) is convergent and there is a Big Bang singularity at $t=0$.

Another result is the following \cite{Seno,proko}: if $H(t_0)>0$ for some $t_0$ and if $R_{\mu \nu}u^\mu u^\nu \ge 0$, then $a(t)$ is vanishing in the past, say $a(0)=0$, then  one has  past-incomplete geodesic.

Following \cite{proko} these results can be extended to future singularities.
Summarizing, in a FLRW spacetime if $a(t) >0$ for all $t$, there are no singularities.
 
\subsection{ The covariant form for the regularity criterion}

In order to formulate the regularity criterion in a covariant form, we work within the metric (\ref{metric2}), and  we introduce the invariant scalar in the reduced spacetime, and thus scalar in the whole spacetime, defined by
\begin{equation}\label{C}
 \Phi(x)=\frac{1-Z(x)}{r^2(x)}\,,
\end{equation}
The regularity criterion then states: a Dynamical Spherically Symmetric spacetime is regular as soon as the invariant quantity $\Phi$ is {\it bounded} for every $x$ of the associated spacetimes.

It should be noted that in GR, the invariant $\Phi$ is related to Misner-Sharp energy by
\begin{equation}\label{CMS}
 \Phi(x)=\frac{2 E_{MS}}{r^3(x)}\,,
\end{equation}
namely, the criterion states that one has a regular DSS spacetime as soon as the Misner-Sharp density is finite everywhere.

First, let us consider a static SS spacetime in the Schwarzschild diagonal gauge, namely 
\begin{equation}\label{sss}
 ds^2 = -b^2(r) f(r)dt^2  + \frac{ dr^2}{f(r)} + r^2 dS^2\,.
\end{equation}
In this gauge, the coordinate $r$ is the  area radius, and one has
\begin{equation}\label{f}
 Z(r) = f(r)\,.
\end{equation}
We recall that the metric represents a BH solution if there exists $r_H>0$ such that $f(r_H)=0$. If we apply the above criterion, we see that  a critical point is $r=0$.  Thus, for small $r$,  one has to require
\begin{equation}\label{f1}
\Phi(r)=\frac{1-f(r)}{r^2}=A+O(r)\,.
\end{equation}
Thus, for spherical horizon, one obtains $ f(r)=1- Ar^2+O(r^3)$, namely the existence of a de Sitter core at the origin. For example, one of the simplest regular BH is the one proposed by Hayward (see also \cite{Is}),
\begin{equation}\label{f}
  f(r)=1-\frac{2m r^2}{r^3+ml^2}\,.
\end{equation}
In this case
\begin{equation}\label{f}
 \Phi(r) = \frac{2m}{r^3+ml^2}\,,
\end{equation}
which is bounded as soon as $l$, Hayward parameter, is not vanishing. Of course, the de Sitter space time in the static patch, obviously statisfies the regularity criterion, since $f(r)=1-H_0^2 r^2$, and $\Phi=H^2_0$.

In the other gauge where the SSS reads
\begin{equation}\label{sssS}
 ds^2 = -dt^2 F(\rho) +  d\rho^2 + r^2(\rho) dS^2\,,
\end{equation}
one has $Z(\rho)=(\frac{d r}{d \rho})^2$. Thus, for small $r$, namely for small $\rho$, one has
\begin{equation}\label{f12}
\Phi(r)=\frac{1-(\frac{d r}{d \rho})^2 }{r^2}=A + O(\rho)\,,
\end{equation}
and one gets the differential equation
\begin{equation}\label{sssSS}
\frac{d r}{d \rho}=\sqrt{1-Ar^2}+O(r^3)\,,
\end{equation}
the solution being
\begin{equation}\label{sssSS1}
 r(\rho)=\frac{\tan(\sqrt{A}\rho)}{\sqrt{A+ \tan^2(\sqrt{A}\rho)}}+O(\rho^4)= \rho-B\rho^3+..\,,
\end{equation}
recovering the  criterion presented in \cite{Seno}.

As final remark for the SSS spacetimes, we observe that the requirement $\Phi(r)=\frac{1-f(r)}{r^2}$ bounded for every $r$ implies that asymptotically $f(r)$ goes at least has $r^2$ for large $r$.

Let us now consider a non-flat FLRW spacetime
\begin{equation}\label{frw}
 ds^2 = -dt^2 + a^2(t) \left( \frac{dr^2}{1-R_0r^2} + r^2 d\Omega^2 \right)\,.
\end{equation}
where $R_0$ is a curvature parameter, namely $R_0$ can be positive, negative or vanishing.
One can compute the invariant $\Phi$ in this case, and with the choice $c_\kappa=1$, one has
\begin{equation}
\Phi=\frac{R_0+\dot a(t)}{a(t)}\,.
\end{equation}
As a consequence, the condition $\Phi$ bounded gives $a(t) \neq 0$ for every $t$. Previously, we have shown that this is equivalent to the geodesics completeness of the related spacetime.

As a simple check, we again consider  the de Sitter space, first in the cosmological flat patch,
\begin{equation}\label{ds0}
 ds^2 = -dt^2 + e^{2H_0} \left(dr^2 + r^2 d\Omega^2 \right)\,,
\end{equation}
and one has $\Phi_0=H_0^2$. If we consider dS  in the complete patch
\begin{equation}\label{ds1}
 ds^2 = -dt^2 + \cosh^2 H_0 t \left( \frac{dr^2}{1-H^2_0r^2} + r^2 d\Omega^2 \right)\,.
\end{equation}
Again the result is $\Phi_0=H_0^2$. In this form, as well known, dS is geodesic complete in the past and future.

\section*{Acknowledgments}

This research has been supported by TIFPA-INFN  within INFN project FLAG. We would like to thank Luciano Vanzo, Max Rinaldi, Lorenzo Sebastiani  and Guido Cognola for useful discussions.


\begin{thebibliography}{9}

\bibitem{LIGO}
	 B.P. Abbott {\it et al.} [LIGO Scientific and Virgo Collaborations],
	{\it Observation of Gravitational Waves from a Binary Black Hole Merger},
	Phys. Rev. Lett. {\bf 116}, 061102 (2016).

\bibitem{LIGO2}
	 B.P. Abbott {\it et al.} [LIGO Scientific and Virgo Collaborations],
	{\it GW170817: Observation of Gravitational Waves from a Binary Neutron Star Inspiral},
	Phys. Rev. Lett. {\bf 119}, 161101 (2017).

\bibitem{LIGO3}
	 B.P. Abbott {\it et al.} [LIGO Scientific and Virgo Collaborations],
	{\it GW170608: Observation of a 19-solar-mass Binary Black Hole Coalescence},
	arXiv:1711.05578 [astro-ph.HE], (2017).
	
\bibitem{LIGO4}
	 B.P. Abbott {\it et al.} [LIGO Scientific and Virgo Collaborations],
	{\it GW170814: A Three-Detector Observation of Gravitational Waves from a Binary Black Hole Coalescence},
	Phys. Rev. Lett. {\bf 119}, 141101 (2017).
	
	\bibitem{LIGO5}
	 B.P. Abbott {\it et al.} [LIGO Scientific and Virgo Collaborations],
	{\it GW170104: Observation of a 50-Solar-Mass Binary Black Hole Coalescence at Redshift 0.2},
	Phys. Rev. Lett., {\bf 118} (22), 221101 (2017).

	\bibitem{LIGO6}
	 B.P. Abbott {\it et al.} [LIGO Scientific and Virgo Collaborations],
	{\it GW151226: Observation of Gravitational Waves from a 22-Solar-Mass Binary Black Hole Coalescence},
	Phys. Rev. Lett. {\bf 116}, 241103 (2016).
	

\bibitem{SingTh1}
	 R. Penrose,
	{\it Gravitational collapse  and space-time singularities},
	Phys. Rev. Lett. {\bf 14}, 57–59 (1965).
	
\bibitem{SingTh2}
	 S. Hawking,
	{\it The occurrence of singularities in cosmology. III. Causality and singularities},
	Proc. Roy. Soc. Lon. A {\bf 300} (1967), 187-201.

\bibitem{SingTh3}
	 S. Hawking and R. Penrose,
	{\it The singularities of gravitational collapse and cosmology},
	Proc. Roy. Soc. Lon. A {\bf 314} (1970), 529-548.


\bibitem{SEMICLASS1}
	 M. Bojowald and A. Skirzewski,
	{\it Effective equations of motion for quantum systems},
	Reviews of Mathematical Physics, {\bf 18}, 713 (2006).

\bibitem{SEMICLASS2}
	 Guillermo Chacon-Acosta, Hector H. Hernandez,
	{\it Effective quantum equations for the semiclassical description of the Hydrogen atom},
	arXiv:1110.3337 [quant-ph], 2011.


\bibitem{AsymSaf1}
	 R. Torres,
	{\it Non-Singular Black Holes, the Cosmological Constant and Asymptotic Safety}, Phys. Rev. D {\bf 95}, 124004 (2017).

\bibitem{AsymSaf2}
	 F. Saueressig, N. Alkofer, G. D'Odorico, F. Vidotto,
	{\it  Black holes in Asymptotically Safe Gravity}, PoS(FFP14) 174 (2014).


\bibitem{LQG}
	 Leonardo Modesto,
	{\it     Loop quantum gravity and black hole singularity},  	arXiv:hep-th/0701239.
	
\bibitem{LQG2}
	 Leonardo Modesto,
	{\it     Space-Time Structure of Loop Quantum Black Hole},  	Int. J. Theor. Phys. {\bf 49}, 1649 (2010).
	
\bibitem{PLANCK2}
	 Marios Christodoulou, Carlo Rovelli, Simone Speziale, Ilya Vilensky
	{\it Realistic Observable in Background-Free Quantum Gravity: the Planck-Star Tunnelling-Time},  	Phys. Rev. D {\bf 94}, 084035 (2016).
	
\bibitem{PLANCK}
	C. Rovelli and F. Vidotto,
	{\it Planck stars}, arXiv:1401.6562 [gr-qc], (2014).
	
	


\bibitem{CONFORMAL1}
	 Cosimo Bambi, Leonardo Modesto, Leslaw Rachwal,
	{\it   Spacetime completeness of non-singular black holes in conformal gravity}, JCAP {\bf 1705}, 003 (2017).

\bibitem{CONFORMAL2}
	 Leonardo Modesto, John W. Moffat, Piero Nicolini,
	{\it   Black holes in an ultraviolet complete quantum gravity}, Phys.Lett. B {\bf 695}, 397-400 (2011).
	
\bibitem{CONFORMAL3}
	 Leonardo Modesto, Leslaw Rachwal,
	{\it    Finite Conformal Quantum Gravity and Nonsingular Spacetimes}, arXiv:1605.04173 [hep-th].
	

\bibitem{Duan}
	Y.S. Duan,
	{\it Generalization of regular solutions of Einstein's gravity equations and Maxwell's equations for point-like charge},
	Soviet Physics JETP, {\bf 27} (6), 756-758 (1954);
	english. transl. by N. Korchagin, [gr-qc/1705.07752].

\bibitem{Bardeen}
	J.M. Bardeen,
	in Conference Proceedings of GR5 (Tbilisi, URSS, 1968), p. 174.


\bibitem{WH1}
	 Gonzalo J. Olmo, D. Rubiera-Garcia,
	{\it  Nonsingular Black Holes in f(R) Theories}, Universe 2015, {\bf 1} (2), 173-185.
	
\bibitem{WH2}
	 Gonzalo J. Olmo, D. Rubiera-Garcia, A. Sanchez-Puente,
	{\it  Classical resolution of black hole singularities via wormholes},  Eur. Phys. J. C {\bf 76}, 143; DOI : 10.1140/epjc/s10052-016-3999-7 (2016).
	
\bibitem{WH3}
	 Cecilia Bejarano, Gonzalo J. Olmo, Diego Rubiera-Garcia,
	{\it   What is a singular black hole beyond General Relativity?}, Phys. Rev. D {\bf 95}, 064043 (2017).

\bibitem{WH4}
	 C. Menchon, Gonzalo J. Olmo, D. Rubiera-Garcia,
	{\it    Nonsingular black holes, wormholes, and de Sitter cores from anisotropic fluids}, Phys. Rev. D {\bf 96}, 104028 (2017).
	




\bibitem{scale}
  E.~Contreras, A.~Rincón, B.~Koch and P.~Bargueño,
  doi:10.1142/S0218271818500323
  arXiv:1711.08400 [gr-qc].

\bibitem{Is}
  E.~Poisson and W.~Israel,
  Class.\ Quant.\ Grav.\  {\bf 5} (1988) L201;
  doi:10.1088/0264-9381/5/12/002.

\bibitem{M}
M.A.  Markov,
{\it Limiting density of matter as a universal law of nature},
JETP Letters {\bf 36}, 265  (1982).

\bibitem{Bronnikov}
	K.A. Bronnikov,
	{\it Regular magnetic black holes and monopoles from nonlinear electrodynamics},
	Phys. Rev. {\bf D 63}, 044005 (2001); doi: 10.1103/PhysRevD.63.044005 [gr-qc/0006014].

\bibitem{Elizalde}
	E. Elizalde, S.R. Hildebrandt
	{\it Family of regular interiors for nonrotating black holes with $T_0^0 = T_1^1$},
	Phys. Rev. {\bf D 65}, 124024 (2002); doi: 10.1103/PhysRevD.65.124024 [gr-qc/0202102v2].

\bibitem{Dymnikova}
	I. Dymnikova,
	{\it Regular electrically charged vacuum structures with de Sitter centre in nonlinear electrodynamics coupled to general relativity},
	Class. Quantum Grav. {\bf 21}, 4417 (2004); doi: 10.1088/0264-9381/21/18/009 [gr-qc/0407072].

\bibitem{GSP}
  I.~H.~Salazar, A.~Garcia and J.~Plebanski,
  J.\ Math.\ Phys.\  {\bf 28} (1987) 2171;
  doi:10.1063/1.527430.

\bibitem{Dymnikova92}
	I.Dymnikova,
	{\it Vacuum nonsingular black hole},
	Gen.Rel.Grav. {\bf 24}, 235 (1992).

\bibitem{ANSS}
	S. Ansoldi, P. Nicolini, A. Smailagic, E. Spallucci,
	{\it Noncommutative geometry inspired charged black holes},
	Phys. Lett. {\bf B 645}, 261 (2007); doi: 10.1016/j.physletb.2006.12.020 [gr-qc/0612035v1].
	
\bibitem{fuzzi}
	Andrea Giugno, Andrea Giusti, Alexis Helou
	{\it  Horizon quantum fuzziness for non-singular black holes}, 
	 arXiv:1711.06209 [gr-qc], (2017).
	 
\bibitem{Maeda}
	G. Kunstatter, H. Maeda, T. Taves,
	{\it New 2D dilaton gravity for nonsingular black holes}, Class. Quant. Grav. {\bf 33}, 105005 (2016).
	
\bibitem{Frolov_3}
	V.P. Frolov,
	{\it Notes on non-singular models of black holes},
	Phys. Rev. {\bf D 94}, no. 10, 104056 (2016); doi:10.1103/PhysRevD.94.104056 [arXiv:1609.01758 [gr-qc]].

\bibitem{NSS}
	P. Nicolini, A. Smailagic, E. Spallucci,
	{\it Noncommutative geometry inspired Schwarzschild black hole},
	Phys. Lett. {\bf B 632}, 547 (2006); doi: 10.1016/j.physletb.2005.11.004 [gr-qc/0510512].

\bibitem{Hayward_1}
	S. Hayward,
	{\it Formation and evaporation of non singular black holes},
	Phys. Rev. Lett. {\bf 96}, 031103 (2006); doi: 10.1103/PhysRevLett.96.031103 [gr-qc/0506126].
	 
\bibitem{Frolov_6}
	V.P. Frolov, M.A. Markov, V.F. Mukhanov,
	{\it Black Holes as Possible Sources of Closed and Semiclosed Worlds},
	Phys. Rev. {\bf D 41}, 383 (1990); doi:10.1103/PhysRevD.41.383.
	 
\bibitem{Frolov_2}
	V.P. Frolov, A. Zelnikov,
	{\it Quantum radiation from an evaporating non-singular black hole}, , Phys. Rev. D 95, 124028 (2017).

\bibitem{Modesto}
	S. Hossenfelder, L. Modesto, I. Prémont-Schwarz,
	{\it A model for non-singular black hole collapse and evaporation},
	Phys. Rev. {\bf D 81}, 044036 (2010); doi: 10.1103/PhysRevD.81.044036 [gr-qc/0912.1823v3].

\bibitem{CMSZ}
	G. Cognola, R. Myrzakulov, L. Sebastiani, S. Zerbini,
	{\it Einstein gravity with Gauss-Bonnet entropic corrections},
	Phys. Rew. {\bf D 88}, 024006 (2013); doi: 10.1103/PhysRevD.88.024006 [gr-qc/1304.1878v2].

\bibitem{Dymnikova_2}
	I. Dymnikova, E. Galaktionov,
	{\it Regular rotating electrically charged black holes and solitons in nonlinear electrodynamics minimally coupled to gravity},
	Class. Quantum Grav. {\bf 32}, 165015 (2015); doi: 10.1088/0264-9381/32/16/165015 [gr-qc/1510.01353v1].

\bibitem{Culetu}
	H. Culetu,
	{\it Microscopic corrections to Schwarzschild spacetime}, (2015) [gr-qc/1508.07570v2].

\bibitem{Horava}
	P. Horava,
	{\it Membranes at quantum criticality},
	JHEP {\bf 0903} 20 (2009); doi: 10.1088/11\%26-6708/2009/03/020 [hep-th/0812.4287v3].

\bibitem{Horava_1}
	P. Horava,
	{\it Quantum gravity at a Lifshitz point},
	Phys. Rev. {\bf D 79} 084008 (2009); doi: 10.1103/PhysRevD.79.084008 [hep-th/0901.3775v2].

\bibitem{KS}
	A. Kehagias, K. Sfetsos,
	{\it The black hole and FRW geometries of non-relativistic gravity},
	Phys. Lett. {\bf B 678}, 123 (2009); doi: 10.1016/j.physletb.2009.06.019 [hep-th/0905.0477v1].

\bibitem{CCO}
	R.G. Cai, L.M. Cao, N. Ohta,
	{\it Black holes in gravity with conformal anomaly and logarithmic term in black hole entropy},
	JHEP {\bf 1004}, 082 (2010); doi: 10.1007/JHEP04(2010)082 [hep-th/0911.4379v2].

\bibitem{Pradhan}
	P. Pradhan,
	{\it Area (or entropy) product formula for a regular black hole}, (2015) [gr-qc/1512.06187].

\bibitem{Ma}
	M.S. Ma,
	{\it Magnetically charged regular black hole in a model of nonlinear electrodynamics},
	Annals Phys. {\bf 362}, 529 (2015); doi:  10.1016/j.aop.2015.08.028 [gr-qc/1509.05580].

\bibitem{Johannsen}
	T. Johannsen,
	{\it Regular black hole metric with three constants of motion},
	Phys. Rev. {\bf D 88}, 044002 (2013); doi: 10.1103/PhysRevD.88.044002 [gr-qc/1501.02809v2].

\bibitem{Rodrigues}
	M.E. Rodrigues, J.C. Fabris, E.L.B. Junior, G.T. Marques
	{\it Generalization of regular black holes in General Relativity to $f(R)$ gravity},
	EPJ {\bf C 76}, 250; doi: 10.1140/epjc/s10052-016-4085-x [gr-qc/1601.00471].

\bibitem{Fan}
	Z.Y. Fan and X. Wang,
	{\it Construction of Regular Black Holes in General Relativity}, arXiv:1610.02636 [gr-qc].

\bibitem{Beato}
	E. Ayon–Beato, A. Garcia,
	{\it Regular black hole in general relativity coupled to nonlinear electrodynamics},
	Phys. Rew. Lett. {\bf 80}, 5056 (1998); doi: 10.1103/PhysRevLett.80.5056 [gr-qc/9911046v1].

\bibitem{Frolov_1}
	V.P. Frolov, A. Zelnikov,
	{\it Quantum radiation from a sandwich black hole},
	Phys. Rev. {\bf D 95}, no. 4, 044042 (2017); doi:10.1103/PhysRevD.95.044042 [arXiv:1612.05319 [hep-th]].

\bibitem{Frolov_4}
	V.P. Frolov,
	{\it Information loss problem and a ’black hole‘ model with a closed apparent horizon},
	JHEP {\bf 1405}, 049 (2014); doi:10.1007/JHEP05(2014)049 [arXiv:1402.5446 [hep-th]].

\bibitem{Frolov_5}
	V.P. Frolov, M.A. Markov, V.F. Mukhanov,
	{\it Through A Black Hole Into A New Universe?},
	Phys. Lett. {\bf B 216}, 272 (1989); doi:10.1016/0370-2693(89)91114-3;

\bibitem{Frolov_7}
	P.A. Bolashenko, V.P. Frolov,
	{\it Certain Properties Of A Nonsingular Model Of A Black Hole}
	in *Markov, M.A. (ED.): The physical effects in the gravitational field of black holes* 205-218.

\bibitem{Vag1} 
  L.~Balart and E.~C.~Vagenas,
  ``Regular black hole metrics and the weak energy condition,''
  Phys.\ Lett.\ B {\bf 730}, 14 (2014).

  \bibitem{Vag2} 
  L.~Balart and E.~C.~Vagenas,
  ``Regular black holes with a nonlinear electrodynamics source,''
  Phys.\ Rev.\ D {\bf 90}, no. 12, 124045 (2014).

      \bibitem{CZ}
	S. Chinaglia, S. Zerbini,
	{A note on singular and non-singular black holes},
	Gen. Relativ. Gravit. {\bf 49}, 75 (2017); doi: 10.1007/s10714-017-2235-6.

 \bibitem{NO17} 
  S.~Nojiri and S.~D.~Odintsov,
  ``Regular Multi-Horizon Black Holes in Modified Gravity with Non-Linear Electrodynamics,''
  Phys.\ Rev.\ D {\bf 96}, no. 10, 104008 (2017).

      \bibitem{ans} 
	S. Ansoldi,
	{\it Spherical black holes with regular center: A Review of existing models including a recent realization with Gaussian sources}, arXiv:0802.0330 [gr-qc].

\bibitem{Sakharov} 
	A. Sakharov,
	{\it Initial stage of an expanding universe and appearance of a nonuniform distribution of matter},
	Sov. Phys. JETP {\bf 22}, 241 (1966).

\bibitem{Mar} 
  M.~A.~Markov,
  ``Problems Of A Perpetually Oscillating Universe,''
  Annals Phys.\  {\bf 155}, 333 (1984).

\bibitem{Stellar}
	T. Roman and P. Bergmann
	{\it Stellar Collapse without Singularities?}, 
	Ph. Rev. D {\bf 28}, (6):1265–1277 (1983).

\bibitem{Barcelo}
	Carlos Barcelo, Raúl Carballo-Rubio, Luis J. Garay
	{\it  Exponential fading to white of black holes in quantum gravity}, Class. Quantum Grav. {\bf 34} 105007 (2017).

\bibitem{Malafarina}
	Daniele Malafarina, 
	{\it Classical collapse to black holes and quantum bounces: A review}, Universe 2017, {\bf 3} (2), 48.

\bibitem{Bambi1}
  C.~Bambi, D.~Malafarina and L.~Modesto,
  Phys.\ Rev.\ D {\bf 88} (2013) 044009
  doi:10.1103/PhysRevD.88.044009
  [arXiv:1305.4790 [gr-qc]].

\bibitem{Bambi2}
	Y. Liu, D. Malafarina, L. Modesto, C. Bambi,
	{\it Singularity avoidance in quantum-inspired inhomogeneous dust collapse}, Phys. Rev. D {\bf 90}, 044040 (2014)

\bibitem{Bambi3}
	C. Bambi, D. Malafarina, L. Modesto,
	{\it Terminating black holes in asymptotically free quantum gravity}, Eur.Phys.J. C {\bf 74}, 2767 (2014).

\bibitem{Bambi4}
	Y. Zhang, Y. Zhu, L. Modesto, C. Bambi,
	{\it Can static regular black holes form from gravitational collapse?}, Eur.Phys.J. C {\bf 75}, 96 (2015).
	
\bibitem{Casadio}
	R. Casadio, S. D.H. Hsu, B. Mirza,
	{\it Asymptotic Safety, Singularities, and Gravitational Collapse}, Phys.Lett. B {695}, 317-319 (2011).

\bibitem{INFOLOSS}
	Samir D. Mathur
	{\it  The information paradox: A pedagogical introduction}, 
	Class.Quant.Grav.{\bf 26}, 224001 (2009).
	
\bibitem{INFOLOSS2}
	Yen Chin Ong
	{\it   Black Hole: The Interior Spacetime}, 
	arXiv:1602.04395 [gr-qc] (2016).
	
\bibitem{INFOLOSS3}
	Sabine Hossenfelder, Lee Smolin
	{\it    Conservative solutions to the black hole information problem}, 
	Phys.Rev.D {\bf 81}, 064009 (2010).


\bibitem{REMN1}
	S.B. Giddings
	{\it  Black Holes and Massive Remnants}, 
	Phys.Rev. D {\bf 46}, 1347-1352 (1992).
	
\bibitem{REMN2}
	S.B. Giddings
	{\it Comments on information loss and remnants}, 
	Phys. Rev. D {\bf 49}, 4078 (1994).
       
\bibitem{REMN3}
	Xavier Calmet
	{\it Virtual Black Holes, Remnants and the Information Paradox}, 
	Class. Quantum Grav. {\bf 32}, 045007 (2015).
	
\bibitem{REMN4}
	Euro Spallucci, Anais Smailagic
	{\it  Regular black holes from semi-classical down to Planckian size}, 
	Int. J. Mod. Phys. D {\bf 26}, n.7, 1730013, (2017).
	
\bibitem{REMN5}
	Pisin Chen, Yen Chin Ong, Dong-han Yeom
	{\it Black Hole Remnants and the Information Loss Paradox}, 
	Physics Reports 1-45, arXiv:1412.8366 [gr-qc] (2015).
	
\bibitem{REMN6}
	Pisin Chen, Ronald J. Adler
	{\it  Black Hole Remnants and Dark Matter}, 
	Nucl.Phys.Proc.Suppl. {\bf 124} 103-106 (2003).
	
\bibitem{REMN7}
	Bernard Carr, Florian Kuhnel
	{\it Primordial black holes as dark matter},
	Phys. Rev. D {\bf 94}, 083504 (2016).
	
\bibitem{REMN8}
	Sebastien Clesse, Juan García-Bellido
	{\it  Seven Hints for Primordial Black Hole Dark Matter},
	 arXiv:1711.10458 [astro-ph.CO] (2017).
	

\bibitem{OTHER0}
  F.~Cunillera and C.~Germani,
  arXiv:1711.01282 [gr-qc].

\bibitem{OTHER1}
	Frederic P. Schuller, Mattias N.R. Wohlfarth
	{\it   Sectional Curvature Bounds in Gravity: Regularisation of the Schwarzschild Singularity}, 
	Nucl.Phys. B {\bf 698}, 319 (2004).

\bibitem{OTHER2}
	Gia Dvali, Cesar Gomez
	{\it  Black Hole's Quantum N-Portrait}, 
	Fortsch.Phys. {\bf 61}, 742-767 arXiv:1112.3359 [hep-th] (2013).
	
\bibitem{OTHER3}
	Pawel O. Mazur, Emil Mottola
	{\it  Gravitational condensate stars: An alternative to black holes}, 
	arXiv:gr-qc/0109035.
	
\bibitem{OTHER4}
	Henrique Gomes, Gabriel Herczeg
	{\it   A Rotating Black Hole Solution for Shape Dynamics}, 
	Class. Quantum Grav. {\bf 31} 175014, (2014).
       
\bibitem{OTHER5}
	Saurya Das
	{\it   Quantum Raychaudhuri equation}, 
	Phys. Rev. D {\bf 89}, 084068 (2014).
	
\bibitem{OTHER6}
	Ovidiu-Cristinel Stoica
	{\it  Schwarzschild Singularity is Semi-Regularizable}, 
	Eur. Phys. J. Plus {\bf 127}, 83 (2012).


\bibitem{Sert}
	O. Sert,
	{\it Regular black hole solutions of the non-minimally coupled $Y(R)F^2$ gravity},
	Journal of Math. Phys. {\bf 57}, 032501 (2016); doi: 10.1063/1.4944428 [gr-qc/1512.01172v2].

\bibitem{Dereli} T. Dereli, Ö. Sert, {\it Non-minimal $ln(R) F^2$ couplings of electromagnetic fields to gravity: static, spherically symmetric solutions}, Eur. Phys. J. {\bf C 71}, 1589 (2011); doi: 10.1140/epjc/s10052-011-1589-2 [gr-qc/1102.3863v1].

\bibitem{Balakin} A.B. Balakin, J.P.S. Lemos, {\it Non-minimal coupling for the gravitational and electromagnetic fields: a general system of equations}, Class. Quantum Grav. {\bf 22}, 1867 (2005); doi: 10.1088/0264-9381/22/9/024 [gr-qc/0503076v2].

\bibitem{Horndeski} G.W. Horndeski, {\it Static spherically symmetric solutions to a system of generalized Einstein-Maxwell field equations}, Phys. Rev. {bf D 17}, 391 (1978); doi: http://dx.doi.org/10.1103/PhysRevD.17.391.

\bibitem{Drummond} I.T. Drummond, S.J. Hathrell, {\it QED vacuum polarization in a background gravitational field and its effect on the velocity of photons}, Phys. Rew. {\bf D 22}, 343 (1980); doi: 10.1103/PhysRevD.22.343.

\bibitem{Zayats}
	A. B. Balakin, A. E. Zayats,
	{\it Non-minimal Wu-Yang monopole}, 
	Phys.Lett. B {\bf 644}, 294-298 (2007).

\bibitem{BLZ}
	A. B. Balakin, J. P. S. Lemos, A. E. Zayats,
	{\it Magnetic black holes and monopoles in a nonminimal Einstein-Yang-Mills theory with a cosmological constant: Exact solutions}, 
	Phys. Rev. D {\bf 93}, 084004 (2016).

\bibitem{NMYM2}
	A. B. Balakin, J. P. S. Lemos, A. E. Zayats,
	{\it Regular nonminimal magnetic black holes in spacetimes with a cosmological constant}, 
	Phys. Rev. D {\bf 93}, 024008 (2016).

\bibitem{DEFELICE}
	Antonio De Felice, Shinji Tsujikawa
	{\it f(R) theories}, 
	Living Rev. Rel. {\bf 13}, 3 (2010).
      
\bibitem{Nicolini}
	P. Nicolini, A. Smailagic, E. Spallucci,
	{\it Reply to arXiv:1704.08516 "A note on singular and non-singular black holes"}, [gr-qc/1705.05359].	

\bibitem{C2017}
  S.~Chinaglia,
  A model of regular black hole satisfying the Weak Energy Condition,
  arXiv:1707.02795 [gr-qc].

\bibitem{Beato_1} E. Ayon-Beato, A. Garcia, {\it The Bardeen model as a non linear magnetic monopole}
Phys. Lett. {\bf B 493}, 149 (2000); doi: 10.1016/S0370-2693(00)01125-4 [gr-qc/0009077].

\bibitem{NBS} M. Novello, S.E.P. Bergliaffa, J.M. Salim, {\it Singularities in general relativity coupled to nonlinear electrodynamics}, Class. Quantum Grav {\bf 17}, 18 (2000); doi: 10.1088/0264-9381/17/18/316 [gr-qc/0003052].
  
 \bibitem{Deser}
	S. Deser, O. Sarioglu, B. Tekin
	{\it  Spherically symmetric solutions of Einstein $+$ non-polynomial gravities}, 
	Gen.Rel.Grav. {\bf 40}, 1-7 (2008).
	
\bibitem{RBHNPG}
	A. Coll\'{e}aux
	{\it Rational regular black holes in non-polynomial gravity}, 
	To appear.
	
\bibitem{LLG1}
	C. Lanczos.,
	{\it A Remarkable Property of the Riemann-Christoffel Tensor in Four Dimensions.}, The Annals of Mathematics, {\bf 39} (4), 842 (1938).
	
\bibitem{Lovelock_2}
	 Lovelock, D.
	{\it  The Einstein tensor and its generalizations}, 
	J. Math. Phys.{\bf 12}, 498-501 (1971).
	
\bibitem{QTG1}
	J. Oliva, S. Ray,
	{\it A new cubic theory of gravity in five dimensions: Black hole, Birkhoff's theorem and C-function}, Class.Quant.Grav.{\bf 27}, 225002 (2010).
	
\bibitem{QTG2}
	R. C. Myers, B. Robinson,
	{\it Black Holes in Quasi-topological Gravity}, JHEP {1008}, 067 (2010).

\bibitem{QTG3}
	J. Oliva, S. Ray,
	{\it Birkhoff's Theorem in Higher Derivative Theories of Gravity}, Class.Quant.Grav. {\bf 28}, 75007 (2011).

\bibitem{QTG4}
	R. A. Hennigar, D. Kubiznak, R. B. Mann,
	{\it Generalized quasi-topological gravity}, Phys. Rev. D {\bf 95}, 104042 (2017).

\bibitem{QTG5}
	Yue-Zhou Li, Hai-Shan Liu, H. Lu,
	{\it Quasi-Topological Ricci Polynomial Gravities}, arxiv, 1708.07198.

\bibitem{QTG6}
  H.~Dykaar, R.~A.~Hennigar and R.~B.~Mann,
  JHEP {\bf 1705} (2017) 045
  doi:10.1007/JHEP05(2017)045
  [arXiv:1703.01633 [hep-th]].

\bibitem{QTG7}
  J.~Ahmed, R.~A.~Hennigar, R.~B.~Mann and M.~Mir,
  JHEP {\bf 1705} (2017) 134
  doi:10.1007/JHEP05(2017)134
  [arXiv:1703.11007 [hep-th]].

\bibitem{QTG8}
	A. Cisterna, L. Guajardo, M. Hassaine and J. Oliva,
	{\it Quintic quasi-topological gravity}, JHEP {1704}, 066 (2017); doi: 10.1007/JHEP04(2017)066.
	
\bibitem{PI}
	E. Poisson, W. Israel,
	{\it  Internal Structure of Black Holes}, 
	Physical Review D, {\bf 41} (6), 1796-1809 (1990); doi: 10.1103/PhysRevD.41.1796.
	
\bibitem{Dilaton1}
	D. Grumiller, W. Kummer, D.V. Vassilevich
	{\it  Dilaton Gravity in Two Dimensions},
	Phys.Rept. {\bf 369}, 327-430 (2002); doi: 10.1016/S0370-1573(02)00267-3.

\bibitem{TavesK}
	T. Taves, G. Kunstatter,
	{\it  Modelling the Evaporation of Non-singular Black Holes}, 
	PhysRev D {\bf 90}, 124062 (2014); doi: 10.1103/PhysRevD.90.124062.

\bibitem{Kunstatter}
	J. Ziprick, G. Kunstatter,
	{\it Quantum Corrected Spherical Collapse: A Phenomenological Framework}, 
	Phys.Rev. D {\bf 82}, 044031 (2010); doi: 10.1103/PhysRevD.82.044031.
	
\bibitem{Designer}
	Gabor Kunstatter, Hideki Maeda, Tim Taves
	{\it Designer black holes from new 2D gravity}, 
	arXiv:1509.04243 [gr-qc].

\bibitem{CCZ17}
	S. Chinaglia, A. Coll\'{e}aux and S. Zerbini,
	{\it A non-polynomial gravity formulation for Loop Quantum Cosmology bounce},
  Galaxies {\bf 5} (2017) no.3,  51; doi: 10.3390/galaxies5030051.
	
\bibitem{bo} 
  M.~Bojowald,
  {\it Absence of singularity in loop quantum cosmology},
  Phys.\ Rev.\ Lett.\  {\bf 86}, 5227 (2001);
  M.~Bojowald,
  {\it Loop quantum cosmology},
  Living Rev.\ Rel.\  {\bf 8}, 11 (2005).
        
\bibitem{Ashtekar}
	A. Ashtekar, T. Pawlowski, P. Singh,
	{\it Quantum nature of the big bang: Improved dynamics},
	Phys. Rev. {\bf D74}, 084003 (2006); [gr-qc/0607039].

\bibitem{Ashtekar_2}
	A. Ashtekar, A. Corichi, P. Singh,
	{\it On the robustness of key features of loop quantum cosmology},
	Phys. Rev. {\bf D77}, 024046 (2008); [0710.3565].
	
\bibitem{Bojo}
	Martin Bojowald
	{\it Consistent Loop Quantum Cosmology}, 
	Class.Quant.Grav. {\bf 26}, 075020 (2009).	
	
\bibitem{Tu1}
J.~Khoury, B.~A.~Ovrut, P.~J.~Steinhardt and N.~Turok,
  {\it The Ekpyrotic universe: Colliding branes and the origin of the hot big bang},
  Phys.\ Rev.\ D {\bf 64} (2001) 123522.

\bibitem{Tsu} 
  S.~Tsujikawa, R.~Brandenberger and F.~Finelli,
  {\it On the construction of nonsingular pre - big bang and ekpyrotic cosmologies and the resulting density perturbations},
  Phys.\ Rev.\ D {\bf 66}, 083513 (2002).

\bibitem{add}
Y.~-S.~Piao, B.~Feng and X.~-m.~Zhang,
  {\it Suppressing CMB quadrupole with a bounce from contracting phase to inflation},
  Phys.\ Rev.\ D {\bf 69}, 103520 (2004);
  Z.~-G.~Liu, Z.~-K.~Guo and Y.~-S.~Piao,
  {\it Obtaining the CMB anomalies with a bounce from the contracting phase to inflation}.
  Phys.\ Rev.\ D {\bf 88}, 063539 (2013).

\bibitem{No}
M.~Novello and S.~E.~P.~Bergliaffa,
  {\it Bouncing Cosmologies},
  Phys.\ Rept.\  {\bf 463} (2008) 127.
  
\bibitem{Bra} 
  R.~Brandenberger,
  {\it Matter Bounce in Horava-Lifshitz Cosmology},
  Phys.\ Rev.\ D {\bf 80}, 043516 (2009).

\bibitem{od}
  K.~Bamba, A.~N.~Makarenko, A.~N.~Myagky, S.~Nojiri and S.~D.~Odintsov,
  JCAP {\bf 1401} (2014) 008
  doi:10.1088/1475-7516/2014/01/008
  [arXiv:1309.3748 [hep-th]].

\bibitem{B}
B.~Xue, D.~Garfinkle, F.~Pretorius and P.~J.~Steinhardt,
  {\it Nonperturbative analysis of the evolution of cosmological perturbations through a nonsingular bounce},
  Phys.\ Rev.\ D {\bf 88} (2013) 083509.
  
\bibitem{lo}
 R.~Myrzakulov and L.~Sebastiani,
  {\it Bounce solutions in viscous fluid cosmology},
  Astrophys.\ Space Sci.\  {\bf 352}, 281 (2014).

\bibitem{staro} 
  B.~Boisseau, H.~Giacomini, D.~Polarski and A.~A.~Starobinsky,
  {\it Bouncing Universes in Scalar-Tensor Gravity Models admitting Negative Potentials},
  JCAP {\bf 1507}, 002 (2015).

\bibitem{Gie} 
  S.~Gielen and N.~Turok,
  {\it Perfect Quantum Cosmological Bounce},
  Phys.\ Rev.\ Lett.\  {\bf 117}, no. 2, 021301 (2016).

\bibitem{I} 
  A.~Ijjas and P.~J.~Steinhardt,
  {\it Fully stable cosmological solutions with a non-singular classical bounce},
  Phys.\ Lett.\ B {\bf 764}, 289 (2017).

\bibitem{qiu} 
  T.~Qiu and Y.~T.~Wang,
  {\it G-Bounce Inflation: Towards Nonsingular Inflation Cosmology with Galileon Field},
  JHEP {\bf 1504}, 130 (2015).

\bibitem{Dz}
	P. Dzierzak, P. Malkiewicz, W. Piechocki,
	{\it Turning Big Bang into Big Bounce: I. Classical Dynamics}, [0907.3436].

\bibitem{Malkiewicz}
	P. Malkiewicz, W. Piechocki,
	{\it Foamy structure of spacetime}, [0907.4647].

\bibitem{Malkiewicz_2}
	P. Malkiewicz, W. Piechocki,
	{\it Turning big bang into big bounce: Quantum dynamics}, CLass. Quant. Grav. {\bf 27}, 225018 (2010); doi: 10.1088/0264-9381/27/22/225018 [gr-qc/0908.4029].

\bibitem{Helling}
	R.C. Helling, G. Policastro,
	{it String quantization: Fock vs. LQG representations}, [hep-th/0409182].

\bibitem{Helling_2}
	R.C. Helling,
	{\it A lesson from the lqg string: Diffeomorphism covariance is enough},
	vol. Proceedings of the Planck Symposium Wrock law. 2009.

\bibitem{Cai}
  Y.~Cai, H.~G.~Li, T.~Qiu and Y.~S.~Piao,
  {\it The Effective Field Theory of nonsingular cosmology: II},
  Eur.\ Phys.\ J.\ C {\bf 77} (2017) no.6,  369
  doi:10.1140/epjc/s10052-017-4938-y
  [arXiv:1701.04330 [gr-qc]].

\bibitem{Cai2}
  Y.~Cai and Y.~S.~Piao,
  {\it A covariant Lagrangian for stable nonsingular bounce},
  JHEP {\bf 1709} (2017) 027
  doi:10.1007/JHEP09(2017)027
  [arXiv:1705.03401 [gr-qc]].

\bibitem{breton} 
  N.~Breton,
  {\it Born-Infeld black hole in the isolated horizon framework},
  Phys.\ Rev.\ D {\bf 67}, 124004 (2003);
  doi:10.1103/PhysRevD.67.124004.
	
\bibitem{MR} M. Rinaldi, {\it Black holes with non-minimal derivative coupling}, Phys. Rew. {\bf D 86}, 084048 (2012); doi: 10.1103/PhysRevD.86.084048 [gr-qc/1208.0103v5].

\bibitem{Hawking_Ellis} S.W. Hawking, G.F.R. Ellis, {\it The large scale structure of spacetime}, Cambridge University Press (1973).
	
\bibitem{Dymnikova_WEC_1} I. Dymnikova, E. Galaktionov, {\it Regular electrically charged vacuum structures with de Sitter center in Nonlinea Electrodynamics coupled to General Relativity },
Class. Quant. Grav. {\bf 21} (2004), 4417-4429; doi: 10.1088/0264-9381/21/18/009.

\bibitem{Dymnikova_WEC_2} I. Dymnikova, E. Galaktionov, {\it Stability of a vacuum nonsingular black hole },
Class. Quant. Grav. {\bf 22}, 2331-2358 (2005).
  
 \bibitem{Dymnikova_1} I. Dymnikova, {\it Spherically symmetric space time with the regular de Sitter center},
Int. J. Mod. Phys. {\bf 1015} (2003).

\bibitem{Solodukhin}
  S.~N.~Solodukhin,
  {\it The Conical singularity and quantum corrections to entropy of black hole},
  Phys.\ Rev.\ D {\bf 51} (1995) 609.

\bibitem{Junior}
  E.~L.~B.~Junior, M.~E.~Rodrigues and M.~J.~S.~Houndjo,
  JCAP {\bf 1510} (2015) 060
  doi:10.1088/1475-7516/2015/10/060
  [arXiv:1503.07857 [gr-qc]].
	
\bibitem{ZerbAim}
	A. Coll\'{e}aux, S. Zerbini,
	{\it Modified gravity models admitting second order equations of motion}, 
	Entropy {\bf 17}, 6643-6662 (2015).
       
\bibitem{gao}
C. Gao
{\it Generalized modified gravity with the second-order acceleration equation    },
  Phys.\ Rev.\ D {\bf 86}, 103512 (2012). 

\bibitem{Deser2}
	S. Deser, A.V. Ryzhov,
	{\it  Curvature invariants of static spherically symmetric geometries}, 
	Class.Quant.Grav. {\bf 22}, 3315-3324 (2005).

\bibitem{AshHay}
	M. C. Ashworth, S. A. Hayward,
	{\it  Boundary Terms and Noether Current of Spherical Black Holes}, 
	Phys.Rev. D {\bf 60}, 084004 (1999).
	
\bibitem{PSC1}
	Palais, Richard S.
	{\it The principle of symmetric criticality}, 
	Commun.Math.Phys. {\bf 69} (1979) no.1, 19-30.

\bibitem{PSC2}
	C. G. Torre,
	{\it Symmetric Criticality in Classical Field Theory}, 
	AIP Conf.Proc. {\bf 1360} (2011) 63-74 arXiv:1011.3429 [math-ph].

\bibitem{DeserFrank}
	S. Deser, J. Franklin,
	{\it  Schwarzschild and Birkhoff a la Weyl}, 
	Am.J.Phys. {\bf 73}, 261-264 (2005).

\bibitem{BRON1}
	Bronstein, M P.,
	{\it Kvantovanie gravitatsionnykh voln (Quantization of Gravitational Waves)},
	Zh.Eksp.Tear.Fiz. {\bf 6}, 195 (1936).
	
\bibitem{BRON2}
	Bronstein, M P.,
	{\it Quantentheorie schwacher Gravitationsfelder},
	Phys.Z.Sowjetunion {\bf 9}, 140-157 (1936).

\bibitem{ROV}
	C. Rovelli, F. Vidotto
	{\it Covariant Loop Quantum Gravity},
	Cambridge University Press, 2014 .
	
\bibitem{LQGMI}
	Eric G. Brown, Robert Mann, and Leonardo Modesto,
	{\it Mass inflation in the loop black hole}, Phys. Rev. D {\bf 84}, 104041 (2011).

\bibitem{Poi} 
  E.~Poisson and W.~Israel,
  ``Inner-horizon instability and mass inflation in black holes,''
  Phys.\ Rev.\ Lett.\  {\bf 63}, 1663 (1989).

\bibitem{MII}
	Andrew J. S. Hamilton, Pedro P. Avelino
	{\it  The physics of the relativistic counter-streaming instability that drives mass inflation inside black holes}, 
	Phys.Rept. {\bf 495}, 1-32 (2010).
	
\bibitem{Mae} 
  H.~Maeda, T.~Torii and T.~Harada,
  {\it Novel Cauchy-horizon instability},
  Phys.\ Rev.\ D {\bf 71}, 064015 (2005).

\bibitem{LOR1}
  T.~De Lorenzo, C.~Pacilio, C.~Rovelli and S.~Speziale,
  Gen.\ Rel.\ Grav.\  {\bf 47} (2015) no.4,  41
  doi:10.1007/s10714-015-1882-8
  [arXiv:1412.6015 [gr-qc]].

\bibitem{LOR2}
	T. De Lorenzo,
	{\it Investigating static and dynamic non- singular black hole}, Master’s thesis, University of Pisa (2014).

\bibitem{Lovelock_1}
	 Lovelock, D.
	{\it  Divergence-free tensorial concomitants}, 
	Aequat. Math. {\bf 4}, 127-138 (1970).

\bibitem{Lovelock_3}
	 Lovelock, D.
	{\it  The four dimensionality of space and the Einstein tensor}, 
	J. Math.Phys.{\bf 13}, 874-876 (1972).
	
\bibitem{Helling_3}
	R.C. Helling,
	{\it Higher curvature counter terms cause the bounce in loop cosmology} (2009); [gr-qc/0912.3011].	

\bibitem{Date}
	Ghanashyam Date, Sandipan Sengupta,
	{\it Effective Actions from Loop Quantum Cosmology:
Correspondence with Higher Curvature Gravity}, 
	Class.Quant.Grav. {\bf 26}, 105002 (2009).

\bibitem{cosmoLove1}
	S. A. Pavluchenko,
	{\it Cosmological dynamics of spatially flat Einstein-Gauss-Bonnet models in various dimensions. Vacuum case}, Phys. Rev. D {\bf 94}, 024046 (2016).
	 
\bibitem{cosmoLove2}
	T. Verwimp,
	{\it On higher dimensional gravity: the Lagrangian, its dimensional reduction and a cosmological model}, Class. Quant. Grav. {\bf 6}, 1655 (1989).

\bibitem{cosmoLove3}
  R.~G.~Cai and S.~P.~Kim,
  JHEP {\bf 0502} (2005) 050
  doi:10.1088/1126-6708/2005/02/050
  [hep-th/0501055].
		
\bibitem{H} 
  H.~Liu, K.~Noui, E.~Wilson-Ewing and D.~Langlois,
  {\it Effective loop quantum cosmology as a higher-derivative scalar-tensor theory},
  arXiv:1703.10812 [gr-qc].
	
\bibitem{Mukhanov_2}
	Ali H. Chamseddine, Viatcheslav Mukhanov
	{\it   Resolving Cosmological Singularities}, 
	JCAP {\bf 1703} no.03, 009 (2017).
	
\bibitem{Mukhanov}
	A.H. Chasmeddine, V. Mukhanov,
	{\it Nonsingular black hole},
	Eur. Phys. J. {\bf C77} no.3, 183 (2017).
	
\bibitem{mi_1}
	Ali H. Chamseddine, Viatcheslav Mukhanov
	{\it   Mimetic Dark Matter}, 
 	JHEP {\bf 1311}, 135 (2013).	
	        
\bibitem{Muk0} 
  A.~H.~Chamseddine, V.~Mukhanov and A.~Vikman,
  {\it ``Cosmology with Mimetic Matter},
  JCAP {\bf 1406}, 017 (2014).

\bibitem{mi_6}
  Y.~Rabochaya and S.~Zerbini,
  {\it A note on a mimetic scalar–tensor cosmological model},
  Eur.\ Phys.\ J.\ C {\bf 76}, no. 2, 85 (2016).
	
\bibitem{mi_2}
  Katrin Hammer, Alexander Vikman
	{\it   Many Faces of Mimetic Gravity}, 
 	arXiv:1512.09118.

\bibitem{mi_3}
  L.~Sebastiani, S.~Vagnozzi and R.~Myrzakulov,
  Adv.\ High Energy Phys.\  {\bf 2017} (2017) 3156915
  doi:10.1155/2017/3156915
  [arXiv:1612.08661 [gr-qc]].
	
\bibitem{mi_4}
	Guido Cognola, Ratbay Myrzakulov, Lorenzo Sebastiani, Sunny Vagnozzi, Sergio Zerbini
	{\it  Covariant Horava-like and mimetic Horndeski gravity: cosmological solutions and perturbations}, 
 	Class. Quant. Grav. {\bf 33}, 22, 225014 (2016).

\bibitem{mi_5}
	 Ratbay Myrzakulov, Lorenzo Sebastiani, Sunny Vagnozzi, Sergio Zerbini
	{\it Static spherically symmetric solutions in mimetic gravity: rotation curves \& wormholes}, 
 	Class. Quant. Grav. {\bf 33}, 12, 125005 (2016).

\bibitem{mi_7}
	 Salvatore Capozziello, Jiro Matsumoto, Shin'ichi Nojiri, Sergei D. Odintsov
	{\it  Dark energy from modified gravity with Lagrange multipliers}, 
	Phys.Lett. B {693},198-208 (2010).

\bibitem{Gali}
  P.~Tretyakov,
  Grav.\ Cosmol.\  {\bf 19} (2013) 288
  doi:10.1134/S0202289313040117
  [arXiv:1302.6343 [gr-qc]].

\bibitem{N3} 
  S.~Nojiri and S.~D.~Odintsov,
 {\it Mimetic $F(R)$ gravity: inflation, dark energy and bounce},
 Mod.\ Phys.\ Lett.\ A {\bf 29}, no. 40, 1450211 (2014).

\bibitem{N4} 
  S.~Nojiri, S.~D.~Odintsov and V.~K.~Oikonomou,
  {\it Ghost-Free $F(R)$ Gravity with Lagrange Multiplier Constraint},
  Phys.\ Lett.\ B {\bf 775}, 44 (2017).

        
\bibitem{Bhar:2017ynp} 
  L.~Herrera and N.~O.~Santos,
  {\it Local anisotropy in self-gravitating systems,}
  Phys.\ Rept.\  {\bf 286}, 53 (1997).

\bibitem{ROTATING}
	 Bobir Toshmatov, Zdeněk Stuchlík, Bobomurat Ahmedov
	{\it  Generic rotating regular black holes in general relativity coupled to nonlinear electrodynamics}, 
	Phys. Rev. D {\bf 95}, 084037 (2017).

\bibitem{N1} 
  S.~Nojiri and S.~D.~Odintsov,
  {\it Unified cosmic history in modified gravity: from F(R) theory to Lorentz non-invariant models},
  Phys.\ Rept.\  {\bf 505}, 59 (2011).

\bibitem{N2} 
  S.~Nojiri, S.~D.~Odintsov and V.~K.~Oikonomou,
  {\it Modified Gravity Theories on a Nutshell: Inflation, Bounce and Late-time Evolution},
  Phys.\ Rept.\  {\bf 692}, 1 (2017).

\bibitem{kanti}
  P.~Kanti, N.~E.~Mavromatos, J.~Rizos, K.~Tamvakis and E.~Winstanley,
  Phys.\ Rev.\ D {\bf 57} (1998) 6255
  doi:10.1103/PhysRevD.57.6255
  [hep-th/9703192].

\bibitem{kanti2}
  G.~Antoniou, A.~Bakopoulos and P.~Kanti,
  arXiv:1711.07431 [hep-th].

\bibitem{kodama}
  H.~Kodama,
  Prog.\ Theor.\ Phys.\  {\bf 63} (1980) 1217.
  doi:10.1143/PTP.63.1217

\bibitem{sean09}
  S.~A.~Hayward, R.~Di Criscienzo, L.~Vanzo, M.~Nadalini and S.~Zerbini,
  Class.\ Quant.\ Grav.\  {\bf 26} (2009) 062001
  doi:10.1088/0264-9381/26/6/062001
  [arXiv:0806.0014 [gr-qc]].

\bibitem{bob09}
  R.~Di Criscienzo, S.~A.~Hayward, M.~Nadalini, L.~Vanzo and S.~Zerbini,
  Class.\ Quant.\ Grav.\  {\bf 27} (2010) 015006
  doi:10.1088/0264-9381/27/1/015006
  [arXiv:0906.1725 [gr-qc]].

\bibitem{noi11}
  L.~Vanzo, G.~Acquaviva and R.~Di Criscienzo,
  Class.\ Quant.\ Grav.\  {\bf 28} (2011) 183001
  doi:10.1088/0264-9381/28/18/183001
  [arXiv:1106.4153 [gr-qc]].
  
\bibitem{Seno}
  J. M.M. Senovilla,
  {\it Singularity theorems and their conseguences }
  Gen. Rel. Grav. 39, 701 (1998). 

\bibitem{Li} 
  M.~Lilley and P.~Peter,
 { \it Bouncing alternatives to inflation},
  Comptes Rendus Physique {\bf 16}, 1038 (2015).

\bibitem{proko} 
H.~H.~Lam and T.~Prokopec,
  ``Singularities in FLRW Spacetimes,''
  arXiv:1606.01147 [gr-qc].

  H.~h.~Lam and T.~Prokopec,
  ``Singularities and Conjugate Points in FLRW Spacetimes,''
  Gen.\ Rel.\ Grav.\  {\bf 49}, no. 10, 133 (2017).
  


	
	
	
  
	
	
	

       
	
	
	
	

       
	
	
	
	

       
	
	
	
	



       

	
	
	
	
	
	

	
	

	
	
	
	
	
	












\end{thebibliography}
\end{document}